%% file: main.tex
\documentclass[conference]{IEEEtran}

\usepackage{paralist}
\usepackage{booktabs}
\usepackage{balance}
\usepackage{amssymb}
\usepackage{graphicx}
\usepackage[most]{tcolorbox}
\usepackage{tabularx}
\usepackage{subfigure}
\usepackage{tikz,pgfplots,pgfplotstable}
\usepackage{mdframed}
\pgfplotsset{compat=1.7}
\usepackage{pifont}
\newcommand{\cmark}{\ding{51}}
\newcommand{\xmark}{\ding{55}}
\usepackage{amsmath}
\usepackage{bm}
\usepackage{multirow}
\usepackage{enumitem}
\usepackage[hyphens]{url}
\usepackage{hyperref}
\hypersetup{%
  pdflang={en-US},
  pdftitle={This PIN Can Be Easily Guessed: Analyzing the Security of Smartphone Unlock PINs},
  pdfauthor={Philipp Markert, Daniel V. Bailey, Maximilian Golla, Markus Dürmuth, Adam J. Aviv},
  pdfsubject={This PIN Can Be Easily Guessed},
  pdfkeywords={Authentication, PIN, Blocklist, Mobile, Smartphone, Security, Usability},
  pdfdisplaydoctitle=true,
  pdfstartview={FitH},
  breaklinks
}
\usepackage{soul}

\definecolor{darkgreen}{RGB}{0,100,0}

\usepackage{multirow}
\def\fourControlAbbr{Con-4}
\def\fourPlaceboAbbr{Pla-4}
\def\fourIosWithAbbr{iOS-4-wC}
\def\fourIosNoAbbr{iOS-4-nC}
\def\fourDataSmallAbbr{DD-4-27}
\def\fourDataLargeAbbr{DD-4-2740}
\def\fourControl{Control-4-digit}
\def\fourPlacebo{Placebo-4-digit}
\def\fourIosWith{iOS-4-digit-wCt}
\def\fourIosNo{iOS-4-digit-nCt}
\def\fourDataSmall{DD-4-digit-27}
\def\fourDataLarge{DD-4-digit-2740}
\def\sixControlAbbr{Con-6}
\def\sixPlaceboAbbr{Pla-6}
\def\sixIosWithAbbr{iOS-6-wC}
\def\sixControl{Control-6-digit}
\def\sixPlacebo{Placebo-6-digit}
\def\sixIosWith{iOS-6-digit-wCt}
\def\fourAmitayAbbr{Amit-4}
\def\fourRockyouAbbr{Rock-4}
\def\sixRockyouAbbr{Rock-6}

\def\fourFirstAbbr{First-4}
\def\sixFirstAbbr{First-6}
\def\fourAmitay{Amitay-4-digit}
\def\fourRockyou{RockYou-4-digit}
\def\sixRockyou{RockYou-6-digit}

\def\fourFirst{First-Choice-4-digit}
\def\sixFirst{First-Choice-6-digit}

\begin{document}

\title{This PIN Can Be Easily Guessed: \\ \huge{Analyzing the Security of Smartphone Unlock PINs} \vspace{-.13cm}}

\author{
\IEEEauthorblockN{Philipp Markert$^{\ast}$}
\IEEEauthorblockA{philipp.markert@rub.de}
\and
\IEEEauthorblockN{Daniel V. Bailey$^{\ast}$}
\IEEEauthorblockA{danbailey@sth.rub.de}
\and
\IEEEauthorblockN{Maximilian Golla$^{\dagger}$}
\IEEEauthorblockA{maximilian.golla@csp.mpg.de}
\and
\IEEEauthorblockN{Markus D\"urmuth$^{\ast}$}
\IEEEauthorblockA{markus.duermuth@rub.de}
\and
\IEEEauthorblockN{Adam J. Aviv$^{\ddagger}$}
\IEEEauthorblockA{aaviv@gwu.edu \\
\hspace{-43.75em} $\ast$ Ruhr University Bochum, $\dagger$ Max Planck Institute for Security and Privacy, $\ddagger$ The George Washington University}
}

\maketitle

\input{00-abstract}

\input{99-tables}
\input{98-figures}

\input{01-intro}
\input{02-related-work}
\input{03-background}
\input{04-user-study}
\input{05-pins}
\input{06-blacklists}
\input{07-conclusion}

\section*{Acknowledgments}
This research was supported by the research training group ``Human Centered Systems Security'' sponsored by the state of North Rhine-Westphalia, Germany, and the German Research Foundation (DFG) within the framework of the Excellence Strategy of the Federal Government and the States -- EXC 2092 CASA -- 390781972. This material is based upon work supported by the National Science Foundation under Grant No. 1845300. Any opinions, findings, and conclusions or recommendations expressed in this material are those of the author(s) and do not necessarily reflect the views of the National Science Foundation.

We also wish to thank Flynn Wolf, Timothy J. Forman, and Leah Flynne for their assistance, and we thank Joseph Bonneau for his revision guidance as well as the feedback from the anonymous reviewers.

\balance
\bibliographystyle{plain}
\bibliography{misc/bibliography}

\clearpage
\nobalance
\appendix
\input{97-appendix}

\end{document}

%% file: 00-abstract.tex
\begin{abstract}
In this paper, we provide the first comprehensive study of user-chosen 4- and 6-digit PINs ($\mathbf{n=1220}$) collected on smartphones with participants being explicitly primed for device unlocking.
We find that against a throttled attacker (with 10, 30, or 100 guesses, matching the smartphone unlock setting), using 6-digit PINs instead of 4-digit PINs provides little to no increase in security, and surprisingly may even decrease security.
We also study the effects of blocklists, where a set of ``easy to guess'' PINs is disallowed during selection.
Two such blocklists are in use today by iOS, for 4-digits (274 PINs) as well as 6-digits (2910 PINs). We extracted both blocklists compared them with four other blocklists, including a small 4-digit (27 PINs), a large 4-digit (2740 PINs), and two placebo blocklists for 4- and 6-digit PINs that always excluded the first-choice PIN.
We find that relatively small blocklists in use today by iOS offer little or no benefit against a throttled guessing attack.
Security gains are only observed when the blocklists are much larger, which in turn comes at the cost of increased user frustration.
Our analysis suggests that a blocklist at about 10\,\% of the PIN space may provide the best balance between usability and security. 
\end{abstract}

%% file: 99-tables.tex
\newcommand{\tabletreatments}[0]{
\begin{table}
\vspace{-.2em}
\caption{Overview of studied treatments.}\label{table:treatments}
\centering
\resizebox{\columnwidth}{!}{
\begin{tabular}{@{}clllrcc@{}} \toprule
   &\textbf{Treatment} & \textbf{Short} & \textbf{Blocklist} & \textbf{Size} &\textbf{Click-thr.} \\
    \midrule
    \multirow{6}{.1em}{\rotatebox{90}{\textbf{4~digits}}} 
   & \fourControl   & \fourControlAbbr   & $-$ & $-$ & $-$ \\
   & \fourPlacebo   & \fourPlaceboAbbr   & First choice & 1 & \xmark \\
   & \fourIosWith   & \fourIosWithAbbr   & iOS 4-digit & 274 & \cmark \\
   & \fourIosNo     & \fourIosNoAbbr     & iOS 4-digit & 274 & \xmark \\
   & \fourDataSmall & \fourDataSmallAbbr & Top Amitay & 27 & \xmark \\
   & \fourDataLarge & \fourDataLargeAbbr & Top Amitay & 2740 & \xmark \\
    \midrule
    \multirow{3}{.1em}{\rotatebox{90}{\textbf{6~digits}}} & \sixControl & \sixControlAbbr & $-$ & $-$ & $-$ \\
  & \sixPlacebo    & \sixPlaceboAbbr    & First choice & 1 & \xmark \\
  & \sixIosWith    & \sixIosWithAbbr    & iOS 6-digit & 2910 & \cmark \\
\bottomrule
\end{tabular}}
\vspace{-.1in}
\end{table}
}

%% file: 98-figures.tex
\newcommand{\figsurveypictures}[0]{
\begin{figure*}[t]
  \centering
\begin{minipage}{0.29\linewidth}
\centering
  \includegraphics[width=0.8\linewidth]{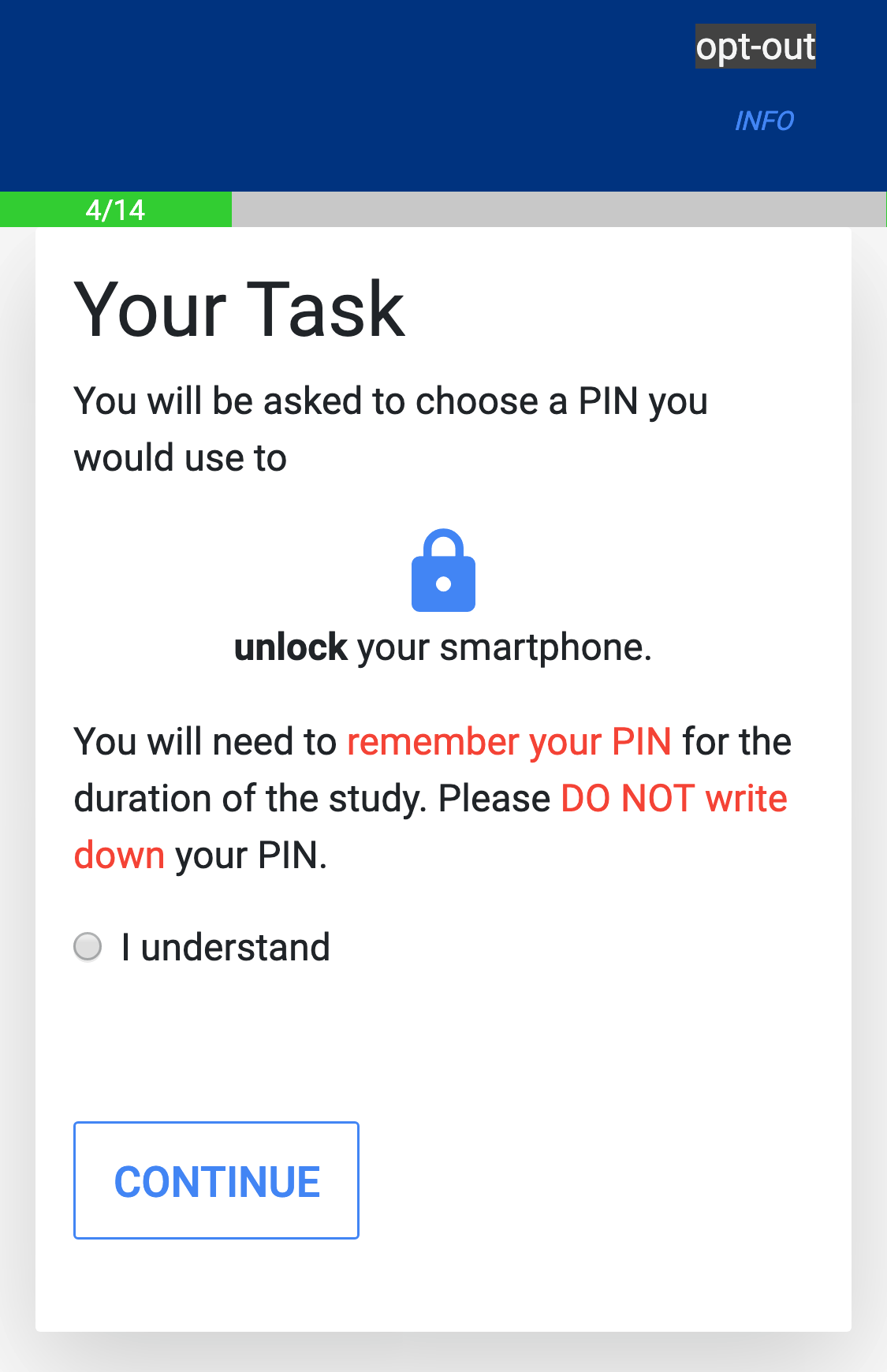}
  \caption{Priming information provided before the participants were asked to create a PIN.}
  \label{fig:priming}
\end{minipage}
\hfill
\begin{minipage}{0.29\linewidth}
\centering
\includegraphics[width=0.8\columnwidth]{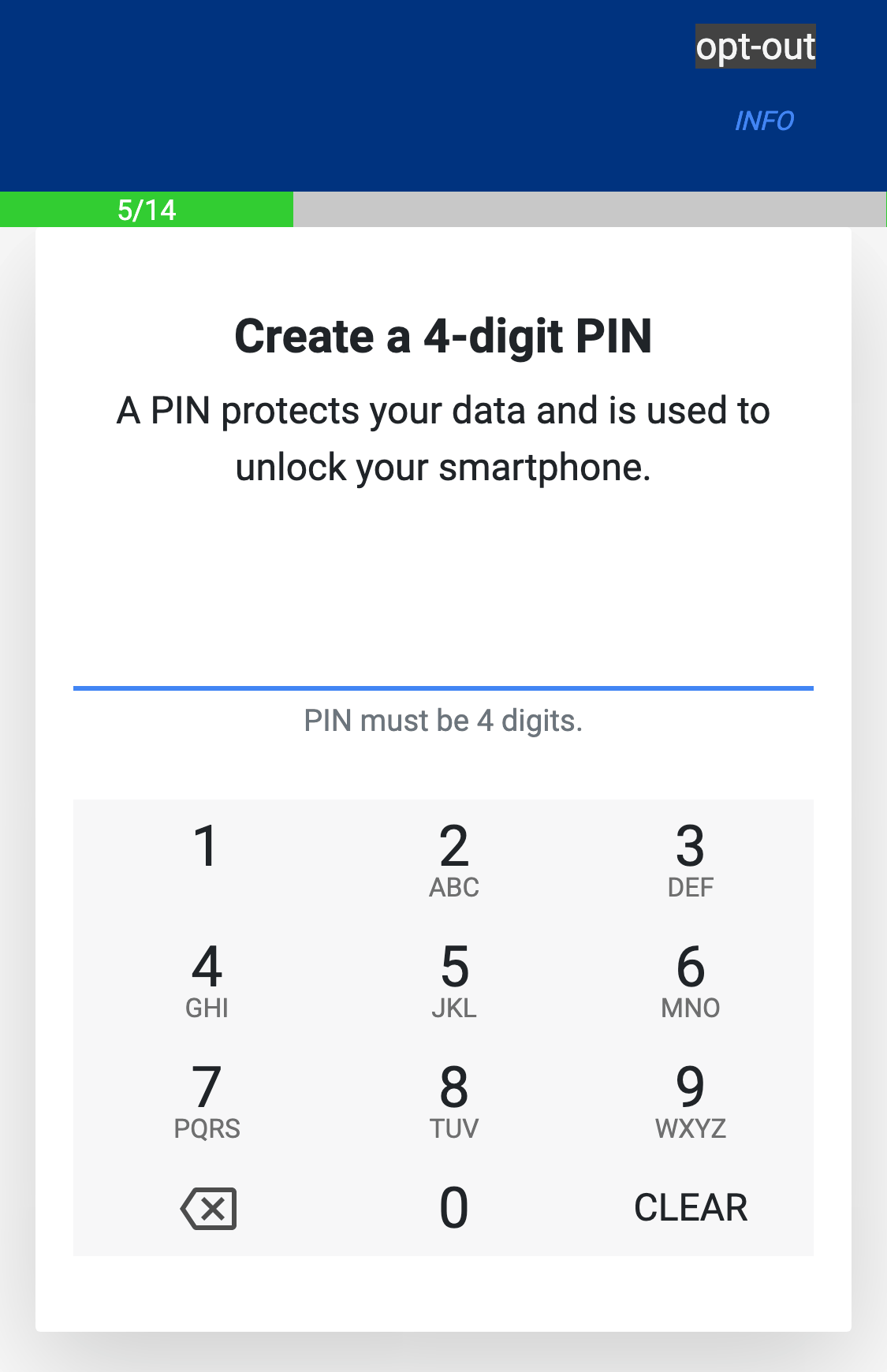}
\caption{The design of the page on which we asked the participants to create a PIN.}
\label{fig:pincreation}
\end{minipage}
\hfill
\begin{minipage}{0.29\linewidth}
\centering
{\includegraphics[width=0.75\columnwidth]{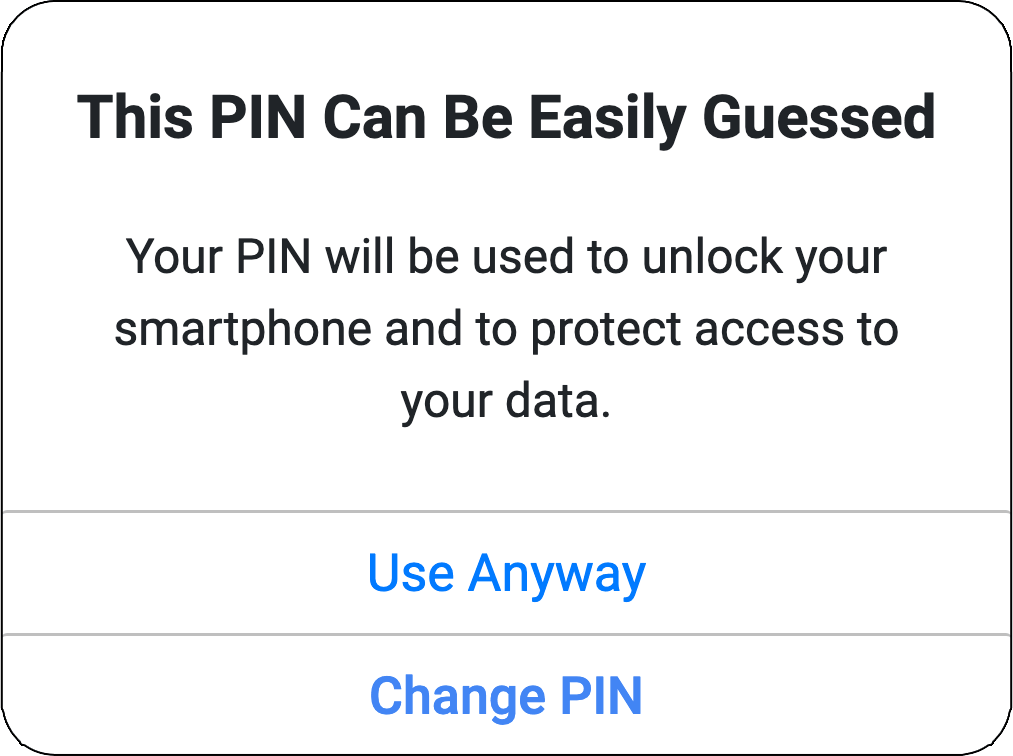}}
\caption{Blocklist warning \textbf{with} the ability to ``click through.''}
\label{fig:feedback:ct}
\smallskip
{\includegraphics[width=0.75\columnwidth]{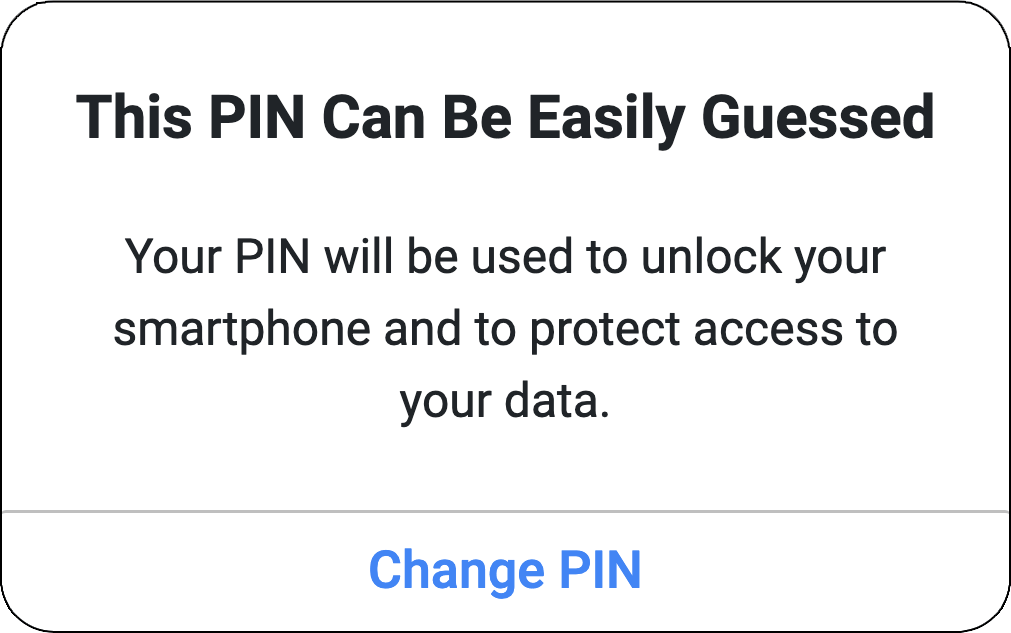}}
\caption{Blocklist warning \textbf{without} the ability to ``click through.''}
\label{fig:feedback:noct}
\vspace{0.2em}
\end{minipage}
\vspace{-1em}
\end{figure*}
}

\newcommand{\figpinextraction}[0]{
\begin{figure}[t]
\centering
    \includegraphics[width=0.9\columnwidth]{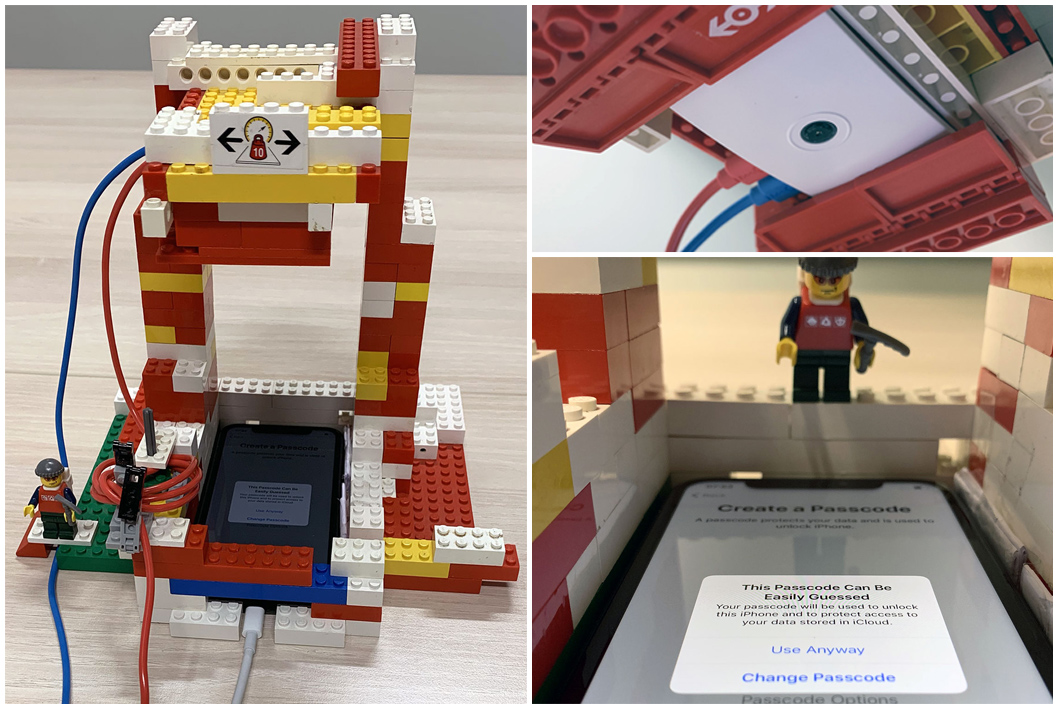}
    \vspace{-.5em}
    \caption{The installation used to extract the iOS blocklists.}
    \label{fig:ios-bl-extraction-1}
    \vspace{-1.1em}
\end{figure}
}

%% file: 01-intro.tex
\section{Introduction}
\label{sec:intro}
We provide the first study focused on the selection of Personal Identification Numbers~(PINs) based on data collected from users specifically primed for the smartphone setting.
While authentication on mobile devices has been studied in several contexts, including patterns~\cite{uellenbeck-13-pattern} and passwords~\cite{melicher-16-smartphone-passwords}, little is known about PINs used for mobile authentication.

Despite the rise of biometrics, such as fingerprint or facial recognition, devices still require PINs, e.g., after a restart or when the biometric fails.
That is because biometric authentication does not replace knowledge-based authentication;
access to a device is still possible with a PIN even when using a biometric.
Moreover, the presence of a biometric may actually lead to a false sense of security when selecting knowledge-based authenticators~\cite{cherapau-15-biometrics-passcodes}.

Our study focuses  on the PINs users choose to unlock their mobile devices.  Previous work on
PINs was primarily focused on  the context of banking, e.g., as part of the \emph{Chip-and-PIN} system~\cite{bonneau-12-pin}
and also mainly relied on the analysis of digit sequences found in leaked text-based password datasets since this data is more readily available~\cite{wang-17-pin}.

Given the sparsity of explicit information about PINs in the context of mobile unlock authentication, we sought to fill this vital knowledge gap by conducting the first study \mbox{($n=1220$)} on the topic
where participants either selected a 4- or 6-digit PIN, the two predominant PIN lengths used for device unlock.
In addition to only allowing participants to complete the study on a smartphone, we also primed our participants specifically for the mobile unlock authentication setting, reminding them that the selected ``PIN protects [their] data and is used to unlock [their] smartphone.''
While our study cannot speak to memorability of selected PINs due to the short time duration, our qualitative feedback suggests that participants took this prompt seriously and selected relevant PINs.

PINs of 4 and 6 digits only provide security when paired with system controls like lockouts and delays that limit offline (or \emph{unthrottled}) guessing.
An unthrottled attacker who can bypass these controls can quickly guess all PIN combinations.
We instead consider a {\em throttled} attacker model to empirically analyze the security of PINs when the system limits the guessing rate.
This is usual in the smartphone-unlocking setting where pauses are enforced after a certain number of wrong guesses in order to slow attacks down.
Guessing is then limited (or throttled) to, e.g., just 10, 30, or 100 attempts in a reasonable time window, such as a few hours.

In such a model, it is essential to prioritize guessing resistance in the first few guesses. Our study found
little benefit to longer 6-digit PINs as compared to 4-digit PINs.
In fact, our participants tend to select more-easily guessed 6-digit PINs when considering the first 40~guesses of an attacker.

As a mechanism for improving PIN selection, we also studied how PINs are affected by blocking.
A blocklist is a set of ``easy to guess'' PINs, which triggers a warning to the user.
Apple iOS devices show the warning {\em ``This PIN Can Be Easily Guessed''} with a choice to {\em ``Use Anyway''} or {\em ``Change PIN.''}
Previous work in text-based passwords has shown that users choose stronger passwords due to a blocklist~\cite{kelley-12-again,shay-15-sugar},
and recent guidance from NIST~\cite{nist-17-sp800-63b} concurs.

To understand selection strategies in the presence of a blocklist,
we conducted a between-subjects comparison of PIN selection using a number of different blocklists. This included one small (27 4-digit~PINs), one large (2740 4-digit~PINs), and two blocklists (274 4-digit PINs and 2910 6-digit PINs) in use today on iOS devices, which we extracted for this purpose.
To determine if the experience of hitting a blocklist or the content of the blocklist itself drives the result, we included a \emph{placebo} blocklist that always excluded the participants' first choice.
Finally, we included both enforcing and non-enforcing blocklists, where participants were able to ``click through'' and ignore the blocklist, the approach taken by iOS.

Despite the popularity of blocklists and the positive impact on textual passwords, our results show that currently employed PIN blocklists are ineffective against
a throttled attacker, in both the enforcing and non-enforcing setting.
This attacker performs nearly as well at guessing 4-digit PINs as if there were no blocklist in use.
To be effective, the blocklist would need to be much larger, leading to higher user frustration. Our results show that a blocklist of about 10\,\% of the PIN space may be able to balance the security and usability needs.

Finally, we collected both quantitative and qualitative feedback from our participants about their PIN selection strategies, perceptions of their PINs in the context of blocklists, and their thoughts about blocking generally.
Overall, we find that despite having mostly negative sentiments about blocklist warnings, participants do perceive the PINs they select under a blocklist as more secure without impacting the memorability and convenience, except in situations of a very large blocklist.

To summarize, we make the following contributions:
\begin{enumerate}[noitemsep,nolistsep]
\item We report on the security of 4\hbox{-} and 6\hbox{-}digit
  PINs as measured for smartphone unlocking, finding that in the throttled
  setting, the benefit of 6-digit PINs is marginal and sometimes
  worse than that of 4-digit PINs.

\item Considering a realistic, throttled attacker model, we show how different
  blocking approaches influence PIN selection process for both security and usability, finding that blocklists in use today
  offer little to no added security.

\item Through quantitative and qualitative feedback, we explore users' perception of security, memorability, and ease-of-use of PIN-based authentication,
  finding that participants perceive that blocking will improve their PINs without impacting usability, except for very large blocklists.

\item We provide guidance for developers on choosing an ap\-pro\-pri\-ate\-ly-sized PIN blocklist
  that can influence the security in the throttled scenario, finding that a 4-digit PIN blocklist needs to be about 10\,\% of the key space to have a noticeable
  impact.

\end{enumerate}
\emph{Note: We responsibly disclosed all our findings to Apple Inc.}

%% file: 02-related-work.tex
\section{Related Work}\label{sec:relatedWork}
Research on PIN authentication for mobile devices is related to the larger area of mobile authentication.
User preferences for different unlock methods for Android devices were studied by Harbach et al.~\cite{harbach-14-hard-lock-life} in 2014.
Since then, PINs have found new uses in encrypting mobile devices~\cite{android-18-masterkey,apple-19-ios-security,newman-19-encrypt-cheap-phones} and biometrics~\cite{cherapau-15-biometrics-passcodes} which require a PIN as part of the keying
material and for fallback authentication when biometrics fail.

The work most closely related to this research is the analysis of PINs in the context of \emph{Chip-and-PIN} systems done by Bonneau et al.~\cite{bonneau-12-pin}, where they considered 4-digit PIN creation strategies for banking customers for use with ATMs/credit cards.
Bonneau et al. identified techniques used for selecting PINs, where choosing (birth) dates/years was the most popular---also true in our setting.
As noted, an attacker can leverage the skewed distribution of PIN choices to improve the guessing strategy.
As a countermeasure, Bonneau et al. proposed the use of a blocklist containing the 100 most popular PINs.
From our analysis, it seems that their suggestion may have formed the basis for Apple iOS's 4-digit blocklist.

Our work differs from Bonneau et al. in two significant ways. Foremost, Bonneau et al. were primarily concerned with payment cards, not smartphone unlock authentication. Second, Bonneau et al. did not collect new PINs
but instead relied on digit sequences found in leaked passwords along with PINs collected without the benefit of a controlled experiment~\cite{amitay-11-iphone-pins}. Our research aims for greater ecological validity by specifically priming users for this task. Our data further suggests that using password leaks may be an imperfect approximation for how users choose PINs for unlock authentication.

Wang et al.~\cite{wang-17-pin} have also analyzed the security of PINs -- in this case without any specific usage context.
They report on comparing 4- and 6-digit PINs created by English and Chinese users.
One counter-intuitive finding is that 6-digit PINs are less resistant to online attacks, despite the key space expansion from 4- to 6-digit PINs.
Our results support the observation that in a rate limited guessing scenario there may actually be no benefit of using 6-digit PINs at all and in certain cases security even decreases.
Yet, Wang et al. used PINs extracted from leaked, text-based password datasets whereas we tend to increase the ecological validity of our results by collecting new PINs specifically primed for mobile authentication and the smartphone form-factor with its standard PIN layout.

Blocklists have been considered in the context of PINs by Kim et al.~\cite{kim-12-pin-policies}.
They tested blocklists for both 4- as well as 6-digit PINs, and concluded that a reasonably-sized blocklist could indeed increase the security.
Kim et al. used \emph{Shannon entropy} and \emph{guessing entropy} as the strength metric and thus only consider an unthrottled, perfect knowledge attacker that will exhaustively guess the PIN space~\cite{bonneau-12-entropy}. This is a questionable attacker model especially given the sparsity of their dataset. Kim et al. compared blocklists representing 2\,\% and 32\,\% of the possible PIN space and found the large blocklist led to lower Shannon-entropy and lower offline guessing-entropy PINs, perhaps due to the composition of Kim et al.'s large blocklist.
In contrast, we show that with a more realistic rate-limited, online attacker, a larger blocklist containing 27.4\,\% of all possible PINs provides a benefit over a smaller one that blocklists only 2.7\,\%, differing from the suggestion of Kim et al. regarding the effect of the size of the blocklist.

Beyond PINs, another common knowledge-based mobile authentication mechanism are Android unlock patterns, whereby a user selects a pattern that connects points on a 3x3 grid.
Uellenbeck et al.~\cite{uellenbeck-13-pattern} showed that user selection of unlock patterns is highly biased, e.g., most patterns start in the upper left corner.
These results have been confirmed by other works~\cite{aviv-15-pattern-bigger,loge-16-pattern-user-choice,zezschwitz-16-pattern-effective-space}.
Most relevant to our study, we  compare the security of mobile unlock PINs to that of patterns and have obtained datasets from related work~\cite{aviv-15-pattern-bigger,loge-16-pattern-user-choice,uellenbeck-13-pattern},~\cite{zezschwitz-16-pattern-effective-space}.

While less common, according to Harbach et al.~\cite{harbach-14-hard-lock-life} and our own measurement (see Table~\ref{tab:device-usage-short}), alphanumeric passwords are another option for users to unlock their mobile devices.
For this reason, we also consider alphanumeric passwords in our comparisons with PINs, as available in leaked, text-based password datasets.
Research has shown that the creation and use of passwords on mobile devices can be cumbersome and users may create weaker passwords than they would do on full-sized keyboards~\cite{greene-14-cant-type-that,melicher-16-smartphone-passwords,schaub-12-passwords-on-smartphones,zezschwitz-14-honey-i-shrunk,yang-14-entry-method-affects}.

%% file: 03-background.tex
\definecolor{structure}{HTML}{03588C} 
\definecolor{note}{HTML}{BF2C47} 

\begin{table}[t]
\caption{Datasets for strength estimations and comparisons.}\label{tab:datasets}
\centering
\resizebox{\columnwidth}{!}{
\begin{tabular}{llrl} \toprule
    \textbf{Kind} & \textbf{Dataset} & \textbf{Samples} & \textbf{Use} \\
    \midrule
    4-digit PINs     & \fourAmitay{}~\cite{amitay-11-iphone-pins}   &      204\,432 & Strength \\
    6-digit PINs     & \sixRockyou{}~\cite{wang-17-pin}   &   2\,758\,490 & Strength \\
    \midrule
    4-digit PINs     & \fourRockyou{}~\cite{wang-17-pin}  &  1\,780\,587 & Comparison \\
    Unlock patterns & ``All'' -- 3x3 patterns~\cite{golla-19-pattern-meter-wip} & 4\,637 & Comparison \\
    Passwords      & LinkedIn~\cite{gosney-16-linkedin-sha1}      & 10\,000 & Comparison \\
    Passwords  & Pwned Passwords v4~\cite{hunt-18-500m} & \emph{Top}~10\,000 & Comparison \\
\bottomrule
\end{tabular}}
\end{table}

\section{Background}
\label{sec:background}

\subsection{Attacker Model}\label{sec:attacker-model}

When studying guessing attackers, there are two primary threat models.
An {\em unthrottled} attacker can guess {\em offline}, indefinitely, until all the secrets are correctly guessed, while a {\em throttled} attacker is limited in the number of guesses, sometimes called an {\em online} attack.
Google's Android and Apple's iOS, the two most popular mobile operating systems, implement real-world rate limiting mechanisms to throttle attackers because otherwise, it would be possible to simply guess all PIN combinations.
In our attacker model, we assume the rate-limiting works as designed, and as such, it is appropriate to  consider a throttled attacker when evaluating security as this best matches the reality of the attacks PINs must sustain for the mobile unlock setting.

The choice of the throttled attack model is further justified when considering mobile devices' {\em trusted execution environments}~(TEE), where the key for device encryption is stored in ``tamper resistant'' hardware and is ``entangled'' with the user's unlock secret~\cite{apple-19-ios-security}.
This forces the attacker to perform decryption (unlock) attempts on the device itself in an online way. Moreover, the TEE is used to throttle the number of decryption attempts tremendously by enforcing rate limiting delays which also survive reboots.\footnote{While there are tools by Cellebrite~\cite{brewster-18-cellebrite} and GrayShift~\cite{brewster-18-graykey} that exploit vulnerabilities in an attempt to escalate guessing to an unthrottled attacker, we consider such attacks out of scope. These exploits are usually bound to a specific device or OS version or can only be run within certain timeframes (e.g., $1$~hour) after the last successful unlock~\cite{welch-18-ios-usb-restricted-mode}.}

An overview of the currently enforced limits is given in Table~\ref{tab:rate-limiting}.
Apple's iOS is very restrictive and only allows up to $10$~guesses~\cite{apple-19-ios-security} before the iPhone disables itself and requires a reset.
Google's Android version~$7$ or newer are less restrictive with a first notable barrier at 30 guesses where the waiting time increases by 10 minutes.
We define the upper bound for a reasonably invested throttled attacker at $100$~guesses when the waiting starts to exceed a time span of 10 hours on Android~\cite{android-19-gatekeeper}, but we also report results for less determined attackers at 10 guesses (30\,s) and 30 guesses (10.5\,m) for Android. The iOS limit is 10 guesses (1.5\,h)~\cite{apple-19-ios-security}.

\begin{table}[t]
    \caption{Rate limiting on mobile operating systems.\vspace{-.05em}}\label{tab:rate-limiting}
  \centering
\resizebox{0.88\columnwidth}{!}{
    \begin{tabular}{r|rr}
    \toprule
    \multicolumn{1}{c|}{\textbf{To Make \emph{n}}} & \multicolumn{2}{c}{\textbf{Accumulated Waiting Time}} \\
    \multicolumn{1}{c|}{\textbf{Guesses}} & \multicolumn{1}{c}{\textbf{Android~7,~8,~9,~10}} & \multicolumn{1}{c}{\textbf{iOS~9, 10, 11, 12,~13}}\\
    \midrule
    1-5 guesses &                    0\,s &       0\,s \\
      6 guesses &                   30\,s &       1\,m 0\,s \\
      7 guesses &                   30\,s &       6\,m 0\,s \\
      8 guesses &                   30\,s &      21\,m 0\,s \\
      9 guesses &                   30\,s &      36\,m 0\,s \\
     10 guesses &                   30\,s & 1\,h 36\,m 0\,s \\
     30 guesses &             10\,m 30\,s & \emph{-} \\
    100 guesses &       10\,h 45\,m 30\,s & \emph{-} \\
    200 guesses & 67\,d  2\,h 45\,m 30\,s & \emph{-} \\
    \bottomrule
    \end{tabular}}
    \vspace{-.6em}
\end{table}

In our attacker model, we assume that the adversary has no background information about the owner of the device or access to other side-channels.
In such a scenario, the best approach for an attacker is to guess the user's PIN in decreasing probability order.
To derive this order, we rely on the best available PIN datasets, which are the \fourAmitay{} and \sixRockyou{} datasets as defined below.
Again, we only consider an {\em un-targeted attacker} who does not have additional information about the victim being attacked. If the attacker is targeted, and is able to use other information and context about the victim, e.g., via shoulder-surfing attack~\cite{schaub-12-passwords-on-smartphones,aviv-17-shoulder-surfing-baseline,aviv-18-comparing-shoulder-surfing} or screen smudges~\cite{aviv-10-smudge}, the attacker would have significant advantages, particularly in guessing 4- vs. 6-digit PINs~\cite{aviv-18-comparing-shoulder-surfing}.

In other parts of this work, we make use of blocklists. In those cases, we consider an attacker that is aware and in possession of the blocklist.  This is because the attacker can crawl the system's blocklist on a sample device, as we have done for this work.
Hence, with knowledge of the blocklist, an informed attacker can improve the guessing strategy by {\em not} guessing known-blocked PINs and instead focusing on common PINs not on the blocklist.

\subsection{Datasets}\label{sec:datasets}

Perhaps the most realistic 4-digit PIN data is from  2011 where Daniel Amitay developed the iOS application ``Big Brother Camera Security''~\cite{amitay-11-iphone-pins}.
The app mimicked a lock screen allowing users to set a 4-digit PIN.
Amitay anonymously and surreptitiously collected 204\,432 4-digit PINs and released them publicly~\cite{amitay-11-iphone-pins}.
While collected in an uncontrolled experiment, we apply the dataset (\fourAmitay{}) when guessing 4-digit PINs, as well as to inform the selection of our ``data-driven'' blocklists.

As there is no similar 6-digit PIN data available to inform the attacker, we rely on 6-digit PINs extracted from password leaks, similar to Bonneau et al.'s~\cite{bonneau-12-pin} and Wang et al.'s~\cite{wang-17-pin} method. PINs are extracted from consecutive sequences of exactly \emph{n}-digits in leaked password data.
For example, if a password contains a sequence of digits of the desired length, this sequence is considered as a PIN (e.g., PW: \texttt{ab3c123456d} $\rightarrow$ PIN: \texttt{123456}, but no 6-digit PINs would be extracted from the sequence \texttt{ab3c1234567d}).

By following this method, we extracted 6-digit PINs from the \emph{RockYou} password
leak, which we refer to as \sixRockyou{} (2\,758\,490 PINs). We also considered
6-digit PINs extracted from other password leaks, such as the
\emph{LinkedIn}~\cite{gosney-16-linkedin-sha1} dataset, but found no marked
differences between the datasets.

To provide more comparison points, we consider a number of other authentication datasets listed in Table~\ref{tab:datasets}.
For example, we use a 3x3 Android unlock pattern dataset described by Golla et al.~\cite{golla-19-pattern-meter-wip}, combining four different datasets~\cite{aviv-15-pattern-bigger,loge-16-pattern-user-choice,uellenbeck-13-pattern,zezschwitz-16-pattern-effective-space}.
It consists of $4637$ patterns with $1635$ of those being unique.
In addition, we use a text-password dataset.
Melicher et al.~\cite{melicher-16-smartphone-passwords} found no difference in strength between passwords created on mobile and traditional devices considering a throttled guessing attacker.
Thus, we use a random sample of 10\,000 passwords from the LinkedIn~\cite{gosney-16-linkedin-sha1} leak and use the \mbox{\emph{Pwned Passwords v4}}~\cite{hunt-18-500m} list to simulate a throttled guessing attacker to estimate the guessing resistance for the sampled LinkedIn passwords as a proxy for mobile text passwords.

\subsection{Extracting the iOS Blocklists}\label{sec:ios-bl-extraction}
As part of our set of blocklists, we also consider a blocklist of ``easily guessed'' 4/6-digit PINs as used in the wild by Apple, which we obtained via brute-force extraction from an iPhone running iOS~$12$. We were able to verify that blocking of PINs is present on iOS~$9$ throughout the latest version iOS~$13$,
and we also discovered that Apple updated their blocklist with the deployment of iOS~$10$ (e.g., the PIN \texttt{101471} is blocked on iOS~$10.3.3$, but is not on iOS~$9.3.5$).

In theory, it is possible to extract the blocklist by reverse engineering iOS, yet, we found a more direct way to determine the blocklist via brute-force:
During device setup, when a PIN is first chosen, there is no throttling.
To test the membership of a PIN, one only needs to enter {\em all} the PINs and observe the presence of the blocklist warning, and then intentionally fail to re-enter the PIN to be able to start over.
We constructed a device to automate this process using a Raspberry~Pi~Zero~W equipped with a Pi~Camera Module~(8MP), as depicted in Figure~\ref{fig:ios-bl-extraction-1}.
The Raspberry~Pi emulates a USB keyboard, which is connected to the iPhone.
After entering a PIN, the camera of the Raspberry~Pi takes a photo of the iPhone screen.
The photo is sent to a remote server, where it is converted to grayscale and thresholded using \emph{OpenCV}.
Subsequently, the presence of the blocklist warning, as depicted in Figure~\ref{fig:feedback:ct}, is detected by extracting the text in the photo using \emph{Tesseract~OCR}.

The extraction of all $10\,000$ 4-digit PINs took $\sim9$~hours.
Testing all $1$~million 6-digit PINs took about $30$~days using two setups in parallel.
To ensure accuracy, we repeated the process for 4-digit PINs multiple times, tested lists of frequent 6-digit PINs, and verified the patterns found in the PINs.
Moreover, we validated all blocked PINs multiple times.
We refer to these two lists as the iOS-4 and iOS-6 blocklists.
\footnote{To foster future research on this topic, we share the described blocklists and the PIN datasets at: \url{https://this-pin-can-be-easily-guessed.github.io}}

In total, the 4-digit blocklist contains $274$~PINs and includes common PINs as well as years from $1956$ to $2015$, but its composition is mostly driven by repetitions such as \texttt{aaaa}, \texttt{abab}, or \texttt{aabb}.
The 6-digit blocklist contains $2910$~PINs and includes common PINs as well as ascending and descending digits (e.g., \texttt{543210}), but its composition is, again, mostly driven by repetitions such as \texttt{aaaaaa}, \texttt{abcabc}, or \texttt{abccba}.
The common PINs blocked by Apple overlap with a 4-digit blocklist suggested by Bonneau et al.~\cite{bonneau-12-pin} in $2012$ and the top 6-digit PINs reported by Wang et al.~\cite{wang-17-pin} in~$2017$.

\figpinextraction{} 

%% file: 04-user-study.tex
\section{User Study}
\label{sec:userStudy}

In this section, we outline the specifics of the treatment
conditions, the user study protocol, and the collected data.  We will also discuss any
limitations of the study as well as ethical considerations.  Please refer to
Appendix~\ref{app:survey} for the specific wording and layouts of the questions.

\subsection{Study Protocol and Design}
We conducted a user study of 4- and 6-digit PINs using Amazon Mechanical
Turk~(MTurk) with $n=1220$~participants over a period of three
weeks.
To mimic the PIN creation process in our browser-based study, participants were restricted to mobile devices by checking the user-agent string.

We applied a 9-treatment, between-subjects study protocol for the PIN
selection criteria, e.g., 4- vs. 6-digit with or without blocking. The
specifics of the treatments are discussed in detail in Section~\ref{sec:treatments}.
At the end of the study, we collected $851$ and $369$ PINs, 4- and 6-digits respectively,  for a total of  $1220$~PINs as our core dataset. These PINs were all selected, confirmed, and recalled. We additionally recorded all intermediate PIN selections, such as what would happen if a selected PIN was {\em not} blocked and the participant did not have to select a different PIN. For more details of different kinds of PINs collected and analyzed, refer to Table~\ref{tab:treatments-results}.

All participants were exposed to a set of
questions and feedback prompts that gauged the security, memorability, and usability of their selected PINs, as well as their attitudes towards blocking events
during PIN selection.

The survey itself consists of 10 parts. Within each part, to avoid ordering effects,
we applied randomization to the order of the questions that may inform later
ones; this information is also available in Appendix~\ref{app:survey}. The parts
of the survey are:
\smallskip

\begin{enumerate}[noitemsep,nolistsep]
\item {\em Informed Consent}: All participants were informed of the procedures
  of the survey and had to provide consent. The informed consent notified
  participants that they would be required to select PINs in different
  treatments, but did not inform them of any details about
  blocking that might be involved in that selection.

\item {\em Agenda}: After being informed, participants were provided additional
  instructions and details in the form of an {\em agenda}. It stated
  the following: ``You will be asked to complete a short survey that
  requires you to select a numeric PIN and then answer some questions about it
  afterwards.  You contribute to research so please answer correctly and as
  detailed as possible.''

\figsurveypictures{}

\item {\em Practice}: Next, participants practiced with the PIN entry
  screen, which mimics typical PIN selection on mobile devices,
  including the ``phoneword'' alphabet on the virtual PIN pad. The purpose of the practice round was to ensure
  that participants were familiar with the interface prior to selecting a
  PIN. There was clear indication during the practice round that this was
  practice and that participants would begin the primary survey afterwards.

\item {\em Priming}: After familiarization and before selection, participants
  were further primed about mobile unlock authentication and PINs using language similar to what iOS and Android use during PIN selection.
   A visual of the priming is in Figure~\ref{fig:priming}.
  A lock icon was used to prime notions of security, and users were reminded that they will need to remember their PIN for the duration of the study without writing it down.
  Participants must click ``I understand'' to continue.
  The qualitative feedback shows that the priming was understood and followed with some participants even stating that they reused their actual PIN.

\item {\em Creation}: The participants then performed the PIN
  creation on the page shown in Figure~\ref{fig:pincreation}.
  The PIN was entered by touching the digits on the virtual PIN pad. As usual, users had to enter the PIN a second time to confirm it was entered correctly.
  Depending on the treatment (see Section~\ref{sec:treatments}), the
  users either selected a 4- or 6-digit PIN and did or did not experience a
  blocklist event.
  In Figure~\ref{fig:feedback:ct} and Figure~\ref{fig:feedback:noct} we depicted the
  two blocklist warnings which either allowed participants to ``click through'' the
  warning (or not). The feedback was copied to directly mimic the wording and layout of a blocklist warning used by Apple since iOS~$12$.

\item {\em Blocklisting Followup}: After creation, we asked participants about
  their attitudes and strategies with blocking. If the participants
  experienced a blocklist event, we referred back to that event in asking
  followup questions. Otherwise, we asked participants to ``imagine'' such an
  experience.  These questions form the heart of
  our qualitative analysis (see Section~\ref{sec:sentiment}).

\item {\em PIN Selection Followup}: We asked a series of questions to gauge
  participants' attitudes towards the PIN they selected with respect to its
  security and usability, where usability was appraised based on ease of entry and
  memorability (see Section~\ref{sec:perception}).
  As part of this questionnaire, we also asked an attention check question.
  We excluded the data of 12 participants because we could not guarantee that they followed our instructions completely.

\item {\em Recall:} On this page, participants were asked to recall their earlier selected
  PIN. Although the two prior parts formed distractor tasks we do not expect that the recall rates measured here speak broadly for the memorability of
  these PINs.
  As expected, nearly all participants could recall their
  selected PIN.

\item {\em Demographics:} In line with best practice~\cite{redmiles-17-best-practices}, we collected the demographics of the
  participants at the very end, including age, gender, IT background,
  and their current mobile unlock authentication.

\item {\em Honesty/Submission:} Finally,
  we asked if the participants
  provided ``honest'' answers to the best of their ability. We informed them that
  they would be paid even if they indicated dishonesty.
Using this information in combination with the attention check described above, we excluded the data of 12 participants to ensure the integrity of our data.
  After affirming
  honesty (or dishonesty), the survey concluded and was submitted.
\end{enumerate}

\subsection{Treatments}
\label{sec:treatments}

We used 9 different treatments: 6~treatments for 4-digit PINs and 3~treatments
for 6-digit PINs. The naming and description of each treatment can be found in
Table~\ref{table:treatments}, as well as the number of participants
(non-overlapping, between-subjects) exposed to each treatment.

\subsubsection{Control Treatments}
For each PIN length, we had a control treatment, \textbf{\fourControl{}} and \textbf{\sixControl{}}, that simply primed
participants for mobile unlock authentication and asked them to select a PIN
without any blocklist interaction. These PINs form the
basis of our 4- and 6-digit mobile-authentication primed PIN dataset. In total,
we have 231 control 4-digit PINs and 127
control 6-digit PINs. We decided to have a larger sample of 4-digit PINs to better validate our methodology compared to other datasets.

We sometimes refer to two datasets, \textbf{\fourFirst{}} and \textbf{\sixFirst{}}.
These combine the control PINs with those chosen by participants from other treatments in their ``first attempt'' before having been subjected to any blocklist.
The \fourFirst{} dataset contains 851 4-digit PINs while \sixFirst{} consists of 369 6-digit PINs.

\subsubsection{Blocklist Treatments}
The remaining treatments considered PIN selection in the presence of a
blocklist.
There are two types of blocklist implementations: {\em enforcing}
and {\em non-enforcing}. An enforcing blocklist does not allow the user to continue as long as the selected PIN is blocked; the
user {\em must} select an unblocked PIN. A non-enforcing blocklist warns the user that the selection is blocked, but the user
can choose to ignore the feedback and proceed anyway.
We describe this treatment as providing the participant
an option to {\em click through}. Otherwise, the treatment uses an enforcing
blocklist. Visuals of the non-enforcing and enforcing feedback can be found in
Figure~\ref{fig:feedback:ct} and~\ref{fig:feedback:noct}, respectively.

\paragraph{Placebo Blocklist}
As we wanted to determine if the experience of hitting a blocklist
or the content of the blocklist itself drive the results, we included a {\em
  placebo} treatment for both 4- and 6-digit PINs (\textbf{\fourPlacebo{}} and
\textbf{\sixPlacebo{}}, respectively). In this treatment, the user's first choice PIN was
blocked, forcing a second choice. As long as the second choice differed from the first, it was accepted.

\paragraph{iOS Blocklist}
For this treatment, we included the blocklists used on Apple's iOS~$12$. The 4-digit
iOS blocklist contains 274 PINs (2.74\,\% of the
available 4-digit PINs), and the 6-digit iOS blocklist contains 2910 PINs
(0.291\,\% of the available 6-digit PINs). These blocklists provide
measurements of real scenarios for users selecting PINs on iOS devices.
As iOS allows users to ``click through'' the blocklist warning and use
their blocked PIN anyway, we implemented our blocking for the iOS
condition in the same way (i.e., conditions \textbf{\fourIosWith{}} and \textbf{\sixIosWith{}}).  To
understand the effect of non-enforcing blocklists, we also had an enforcing
version of the iOS blocklist for 4-digits (\textbf{\fourIosNo{}}).

\paragraph{\mbox{Data-Driven Blocklists}}
We considered two 4-digit block\-lists that are significantly (10x) smaller (27~PINs) and (10x) larger (2740~PINs) than the iOS blocklist.
The blocklists were constructed using the 27 and 2740 most frequently occurring PINs in the \fourAmitay{} dataset, and we refer to them as \textbf{\fourDataSmall{}} and \textbf{\fourDataLarge{}}.
When comparing these two data-driven blocklists and the one used in iOS, it can be seen that they have different compositions.
While 22, i.e., 82\,\% of the PINs contained in \fourDataSmall{} are blocked in iOS, there are also 5 PINs which are not.
Surprisingly, these PINs correspond to simple patterns like \texttt{0852} which is a bottom-up pattern across the PIN pad or \texttt{1379}, the four corners of the pad chosen in a left-to-right manner.
Similar observations can be made when comparing the iOS and the large \fourDataLargeAbbr{} blocklist.
Of 274 PINs which are rejected by iOS, 258, i.e., 92\,\%, are also blocked by our large data-driven blocklist.
The remaining 16 PINs all follow the same repetitive \texttt{aabb} scheme, e.g., \texttt{0033}, \texttt{4433}, or \texttt{9955}.
Interestingly, only one of those PINs, \texttt{9933}, was selected in our study which shows that double repetitions are presumably not as common as Apple expects.

\tabletreatments{}

\subsection{Recruitment and Demographics}
Using Amazon's Mechanical Turk (MTurk), we recruited a total of 1452~participants.
After excluding a portion due to invalid responses to attention tests or survey errors, we had 1220~participants remaining.
We required our participants to be 18~years or older, reside in the US (as checked by MTurk), and have at least an 85\,\% approval rate on MTurk.
The IRB approval required focusing on participants residing in the US, but there may be a secondary benefit to this:
US residents often do not have {\em chip-and-PIN} credit cards (although, they do use 4-digit ATM PINs), in contrast to residents in Europe or Asia, and thus may associate PIN selection more strongly with mobile device locking.
In any case, participants were explicitly primed for the mobile device unlock setting.
Participants indicated they understood this instruction, and their qualitative responses confirm that this was indeed the case.

We also reviewed all of the participants' responses for consistency, including answers to attention check questions, the honesty question, and speed of entry. We removed 12
who provided inconsistent data but did not ``reject'' any participants on Amazon Mechanical Turk.
Participants were compensated with \$\,1~(USD) for completion; the survey took on average 5~minutes for an hourly rate of \$\,12.

\paragraph{Demographics and Background}
As typical on MTurk, our sample is relatively young and better
educated than the general US population. Of the participants, 619 identified as
male (51\%) while 590 (48\%) identified as female (1\,\% identified as other or
preferred not to say), and the plurality of our participants were between 25 and 34 years old (47\,\%).
Most participants had some college (23\,\%) or a
bachelor's degree (39\,\%), and few (12\,\%) had a master's or doctoral degree. While 26\,\%
described having a technical background, 71\,\% described not having one.
We have the full details of the demographics responses in Appendix~\ref{app:demographics} in Table~\ref{tab:demographics}.

\paragraph{Smartphone OS}
We asked participants which operating system
they use on their primary
smartphone. Slightly more than half, 698 (57\,\%), of the participants were
Android users, while 506 (42\,\%) were iOS users. We collected browser user-agent strings
during the survey, and confirmed similar breakdowns, suggesting most
participants used their primary smartphone to take the survey. A detailed
breakdown can be found in the Appendix~\ref{app:device-usage} in Table~\ref{tab:device-usage}.

\paragraph{Unlock Schemes Usage}
As we focus on mobile authentication, we were interested in learning about the kind of mobile authentication our participants use, recalling both biometric and knowledge-based authentication may be in use on a single device.
We first asked if a biometric was used and then asked what authentication participants use instead or as a backup for the biometric, e.g., when it fails.
While Table~\ref{tab:device-usage-short} shows a compressed description, a detailed breakdown can be found in the Appendix~\ref{app:device-usage} in Table~\ref{tab:device-usage}.
For knowledge-based authenticators, considered here, PINs are the most common: 44\,\% described using a 4-digit PIN, 20\,\% using a 6-digit PIN, and 3\,\% using a PIN longer than 6 digits.
The second most common form of knowledge-based authentication are Android unlock patterns at 14\,\%, and 44~participants (or 4\,\%) reported using an alphanumeric
password.
In our study, 140~participants reported not using any locking method.

\begin{table}
    \caption{Usage of mobile unlock authentication schemes.}\label{tab:device-usage-short}
    \vspace{-.5em}
  \centering
\resizebox{\columnwidth}{!}{
  \footnotesize
    \begin{tabular}{rrr|rrr}
    \toprule
    \textbf{Primary Scheme} & \textbf{No.} &  \multicolumn{1}{c|}{\textbf{\%}} & \textbf{Secondary Scheme} & \textbf{No.} & \multicolumn{1}{c}{\textbf{\%}} \\
    \midrule
     \multirow{4}{*}{Fingerprint} & \multirow{4}{*}{573} & \multirow{4}{*}{47\,\%} & 4-digit PIN & 285 & 50\,\% \\
     &&& 6-digit PIN & 148 & 26\,\% \\
     &&& Pattern & 84 & 14\,\% \\
     &&& Other & 56 & 10\,\% \\
     \midrule
     \multirow{4}{*}{Face} & \multirow{4}{*}{162} & \multirow{4}{*}{13\,\%} & 4-digit PIN & 82 & 51\,\% \\
     &&& 6-digit PIN & 50 & 31\,\% \\
     &&& Pattern & 15 & 9\,\% \\
     &&& Other & 15 & 9\,\% \\
     \midrule
     \multirow{4}{*}{Other Biometric} & \multirow{4}{*}{24} & \multirow{4}{*}{2\,\%} & 4-digit PIN & 5 & 21\,\% \\
     &&& 6-digit PIN & 3 & 13\,\% \\
     &&& Pattern & 14 & 58\,\%  \\
     &&& Other & 2 & 8\,\% \\
     \midrule
     4-digit PIN & 165 & 14\,\% \\
     6-digit PIN & 37 & 3\,\% \\
     Pattern & 60 & 5\,\% & \multicolumn{3}{c}{\textit{No secondary scheme used.}} \\
     Other & 59 & 5\,\% \\
     None & 140 & 11\,\% \\
    \bottomrule
    \end{tabular}}
    \vspace{-2em}
\end{table}

\subsection{Ethical Considerations}
All of the survey material and protocol was approved by our Institutional Review
Board (IRB). Beyond meeting the approval of our institution, we worked to uphold
the ethical principles outlined in the Menlo Report~\cite{dhs-12-menlo-report}.

In practicing {\em respect for persons} and {\em justice}, beyond
informing and getting consent, we also sought to compensate participants fairly
at least at the minimum wage of the municipality where the oversight was
performed. Since some of our treatments may frustrate
participants, e.g., where the blocklist was quite large (\fourDataLarge{}), we also
compensated those who returned the survey and notified
us of their frustration.

Additionally, as we are dealing with authentication information, we evaluated
the ethics of collecting PINs and distributing blocklists in terms of {\em beneficence}.
With respect to collecting PINs, there is risk in that
participants may (and likely will) expose PINs used in actual authentication. However, there is limited to no risk in that exposure due to the fact that PINs are not linked to participants and thus cannot be used in a targeted attack. A targeted attack would need proximity and awareness of the victim, of which, neither is the case for this study. Meanwhile, the benefit of the research is high in that the goal of this research is to improve the security of mobile authentication.
Similarly, distributing
blocklists increases social good and scientific understanding with minimal risk
as a determined attacker likely already has access to this material.

Finally, we have described our procedures transparently and make our methods
available when considering {\em respect for law and public interest}. We also do
not access any information that is not already publicly available.

\subsection{Limitations}
There are a number of limitations in this study.
Foremost among them is the fact that the participant sample is skewed towards mostly younger users residing in the US.
However, as we described previously, there may be some benefit to studying PINs from US residents as they are less familiar with \textit{chip-and-PIN} systems and may be more likely to associate PINs directly with mobile unlocking.
We argue that our sample provides realizable and generalizable results regarding  the larger ecosystem of PIN selection for mobile authentication.
Further research would be needed to understand how more
age-diverse~\cite{qiu-19-age-effects} and location-diverse populations select PINs.

Another limitation of the survey is that we are asking participants to select PINs while primed for mobile authentication and there is a risk that participants do not act the same way in the wild. We note that similar priming is used in the authentication literature for both text-based passwords for desktop~\cite{ur-15-added-at-the-end, ur-17-data-driven-pm} and mobile settings~\cite{melicher-16-smartphone-passwords}, and these results generalize when compared to passwords from leaked password datasets~\cite{ur-15-pgs}. We have similar results here. When compared to the most realistic dataset previously available, \fourAmitay{}, the most common 4-digit PINs collected in our study are also present in similar distributions to Amitay~\cite{amitay-11-iphone-pins}. Also, in analyzing the qualitative data, a number of participants noted that they used their real unlock PINs.

While this presents strong evidence of the effectiveness of mobile unlock priming, we, unfortunately, do not have any true comparison points, like what is available for text-based passwords.
There is no obvious analog to the kinds of attacks that have exposed millions of text-based passwords that would similarly leak millions of mobile unlock PINs.
Given the available evidence, we argue
that collecting PINs primed for mobile unlock authentication provides a reasonable
approximation for how users choose PINs in the wild.

Due to the short, online nature of our study, we are limited in what we can
conclude about the memorability of the PINs. The entirety
of the study is only around 5~minutes, while mobile authentication PINs are used
for indefinite periods, and likely carried from one device to the next. There
are clear differences in these cases, and while we report on the
recall rates within the context of the study, these results do not generalize.

Finally, we limited the warning messaging used when a blocklist event
occurred. We made this choice based on evaluating the messaging as used by iOS,
but there is a long line of research in appropriate
security messaging~\cite{sunshine-09-crying-wolf, akhawe-13-alice, felt-15-ssl-warnings, golla-18-what-was}. We do not wish to make claims about the quality of this messaging, and a limitation of this study (and an area of future
work) is to understand how
messaging affects changing
strategies and click-through rates. 

%% file: 05-pins.tex
\section{PIN Selection on Smartphones}
\label{sec:pins}
In the following section, we discuss the security of both 4- and 6-digit PINs.
Unless otherwise stated, our analyzed dataset consists of the PINs entered before any blocklist warning in Step~(5) of the study. These so-called ``first choice'' PINs (cf. Table~\ref{tab:treatments-results}) are unaffected by the blocklists.

\input{05-1-strength}
\input{05-2-selection}

%% file: 05-1-strength.tex
\subsection{Strength of 4- and 6-digit PINs}

\paragraph{Entropy-Based Strength Metrics}
We analyzed PINs in terms of their mathematical metrics for guessing resistance based on entropy estimations.
For this, we consider a \emph{perfect knowledge} attacker who always guesses correctly (in perfect order) as described by Bonneau et al.~\cite{bonneau-12-entropy}.
The advantage of such an entropy estimation approach is that it always models a best-case attacker and does not introduce bias from a specific guessing approach.
Our results are given in Table~\ref{tab:entropy}.

We report the $\beta$-success-rate, which measures the expected guessing success for a throttled adversary limited to $\beta$-guesses per account (e.g., $\lambda_{3}$ = $3$~guesses).
Moreover, we provide the Min-entropy $H_{\infty}$ as a lower bound estimate that solely relies on the frequency of the most common PIN (\texttt{1234}, \texttt{123456}).
Finally, we present the partial guessing entropy ($\alpha$-guesswork) $G_{\alpha}$, which provides an estimate for an unthrottled attacker trying to guess a fraction $\alpha$ of all PINs in the dataset.
In three cases, the calculation of $\widetilde{G}_{0.2}$ is based on PINs occurring only once, due to the small size of the datasets.
This constraint would result in inaccurate guessing-entropy values which is why they are not reported.

For a fair comparison among the datasets which all differ in size, we downsampled \fourFirstAbbr{}, \fourAmitayAbbr{}, \fourRockyouAbbr, and \sixRockyouAbbr{} to the size of the smallest dataset \sixFirstAbbr{} (369 PINs) in our calculations.
We repeated this process 500~times, removed outliers using Tukey fences with $k=1.5$. In Table~\ref{tab:entropy} we report the median values.

The low Min-entropy of the \sixRockyouAbbr{} dataset is due to the fact that the PIN \texttt{123456} is over-represented.
It is $21\times$ more frequent than the second-most popular PIN.
In contrast, the most common 4-digit PIN occurs only $1.7\times$ more often, leading to a lower $H_{\infty}$ value.

Overall, the PINs we collected, specifically primed for mobile authentication, have different (and \emph{stronger}) strength estimations than PINs derived from leaked text-based password datasets studied in the previous work. This is true for both the 4- and 6-digit PINs, which supports our motivation for conducting studies that collect PINs directly.

\paragraph{Guess Number-Driven Strength Estimates}
Next, we estimate the security of the PINs in regard to real-world guessing attacks.
For this, we consider an attacker as described in Section~\ref{sec:attacker-model}.
Our attacker guesses PINs in decreasing probability order based on the \fourAmitayAbbr{}, \fourRockyouAbbr{}, and \sixRockyouAbbr{} datasets.
When two or more PINs share the same frequency, i.e., it is not possible to directly determine a guessing order, Golla et al.~\cite{golla-18-psm} suggests ordering those PINs using a Markov model.
We trained our model on the bi-grams (4-digit PINs) or tri-grams (6-digit PINs) of the respective attacking datasets which simulates the attacker with the highest success rate for each case without overfitting the problem.

\begin{table}[t]
\vspace{-.15em}
\caption{Guessing difficulty for a perfect-knowledge attacker.}\label{tab:entropy}
    \centering
    \resizebox{1.0\columnwidth}{!}{\begin{tabular}{l|rrc|cccc}
    \toprule
    \multicolumn{1}{c|}{} & \multicolumn{3}{c|}{\textbf{Online Guessing (Success \%)}} & \multicolumn{4}{c}{\textbf{Offline Guessing (bits)}}\\
    \multicolumn{1}{l|}{\textbf{Dataset}} & \multicolumn{1}{c}{\textbf{$\boldsymbol{\lambda_{3}}$}} & \multicolumn{1}{c}{$\boldsymbol{\lambda_{10}}$} & $\boldsymbol{\lambda_{30}}$ & \textbf{$\boldsymbol{H_{\infty}}$} & $\boldsymbol{\widetilde{G}_{0.05}}$ & $\boldsymbol{\widetilde{G}_{0.1}}$ & $\boldsymbol{\widetilde{G}_{0.2}}$\\
    \midrule
    \fourFirstAbbr{}\textsuperscript{\scriptsize{$\dagger$}} & 3.79\,\%  & 7.86\,\% & 16.80\,\% & 5.72 & 6.60 & 7.11  & -\textsuperscript{\scriptsize{$\star$}} \\
    \fourAmitayAbbr{}\textsuperscript{\scriptsize{$\dagger$}}  & 9.49\,\%  & 16.26\,\% & 26.29\,\% & 4.53  & 4.74 & 5.16  & 6.33  \\
    \fourRockyouAbbr{}\textsuperscript{\scriptsize{$\dagger$}}  & 8.67\,\%  & 18.70\,\% & 32.79\,\% & 4.72  & 4.94 & 5.23  & 5.81  \\
    \midrule
    \sixFirstAbbr{}  & 6.23\,\%  & 10.30\,\% & 15.72\,\% & 4.53 & 5.19 & 6.57  & -\textsuperscript{\scriptsize{$\star$}} \\
    \sixRockyouAbbr{}\textsuperscript{\scriptsize{$\dagger$}}  & 13.28\,\%  & 16.53\,\% & 21.95\,\% & 3.10  & 3.10 & 3.07 &  -\textsuperscript{\scriptsize{$\star$}} \\
    \bottomrule
    \end{tabular}}
    \begin{flushleft}
    \vspace{.3em}
    {\scriptsize{$\dagger$: For a fair comparison we downsampled the datasets to the size of \sixFirstAbbr{}~(369~PINs).\\}}
    \vspace{.3em}
    {\scriptsize{$\star$: We omit entries which are not sufficiently supported by the underlying data.}}
  \end{flushleft}
\end{table}

\begin{figure}[t]
\centering
\subfigure[Guessing performance against mobile authentication systems based on the number of guesses.]{
  \includegraphics[width=0.93\columnwidth]{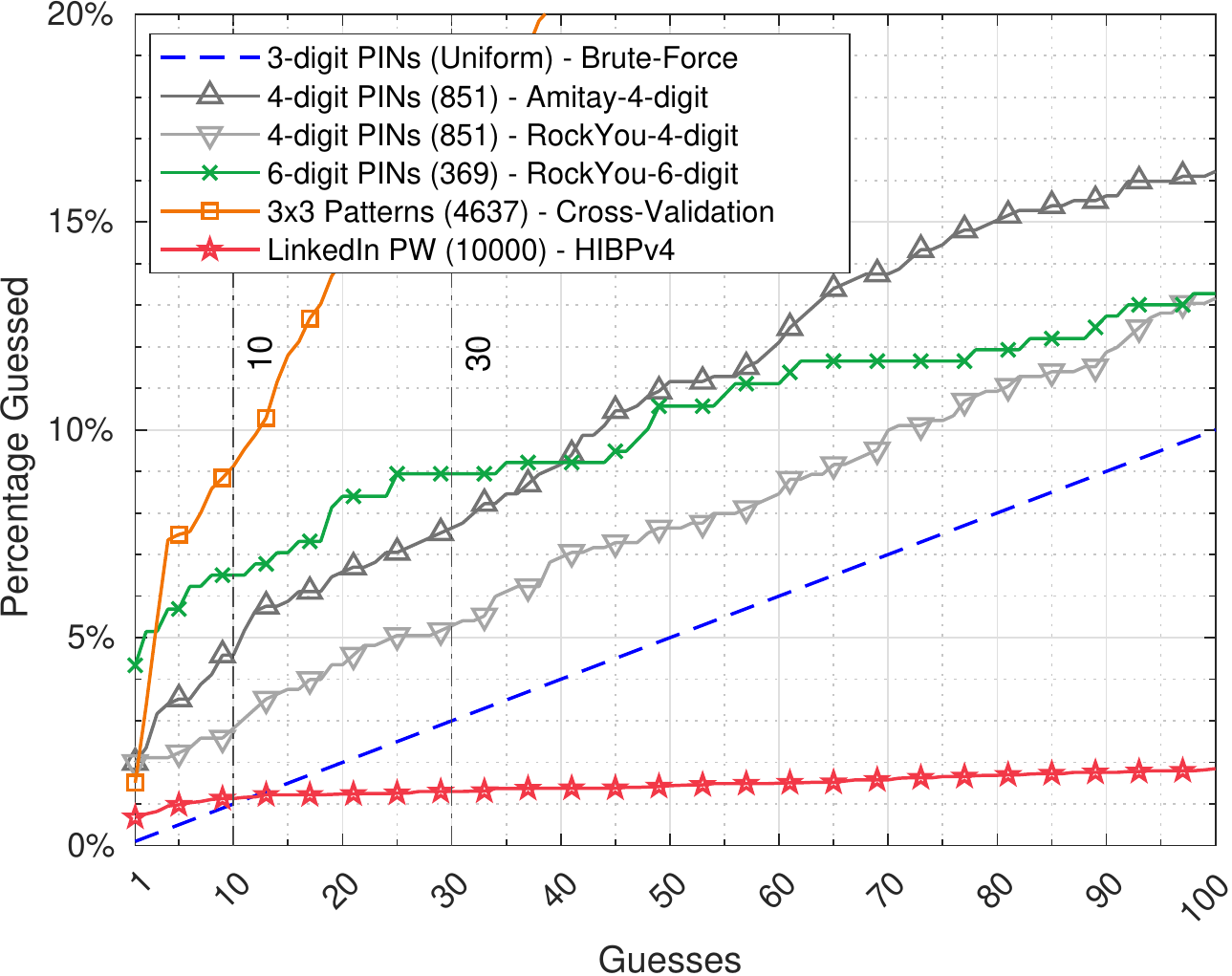}
    \label{fig:comparison-100}
}
\subfigure[Guessing performance against 4- and 6-digit PINs on Android and iOS based on the required time. For 4-digit PINs, we only show the success rate of an attack with \fourAmitayAbbr{} as it outperforms \fourRockyouAbbr{} (cf. Figure~\ref{fig:comparison-100}).]{
    \includegraphics[width=0.93\columnwidth]{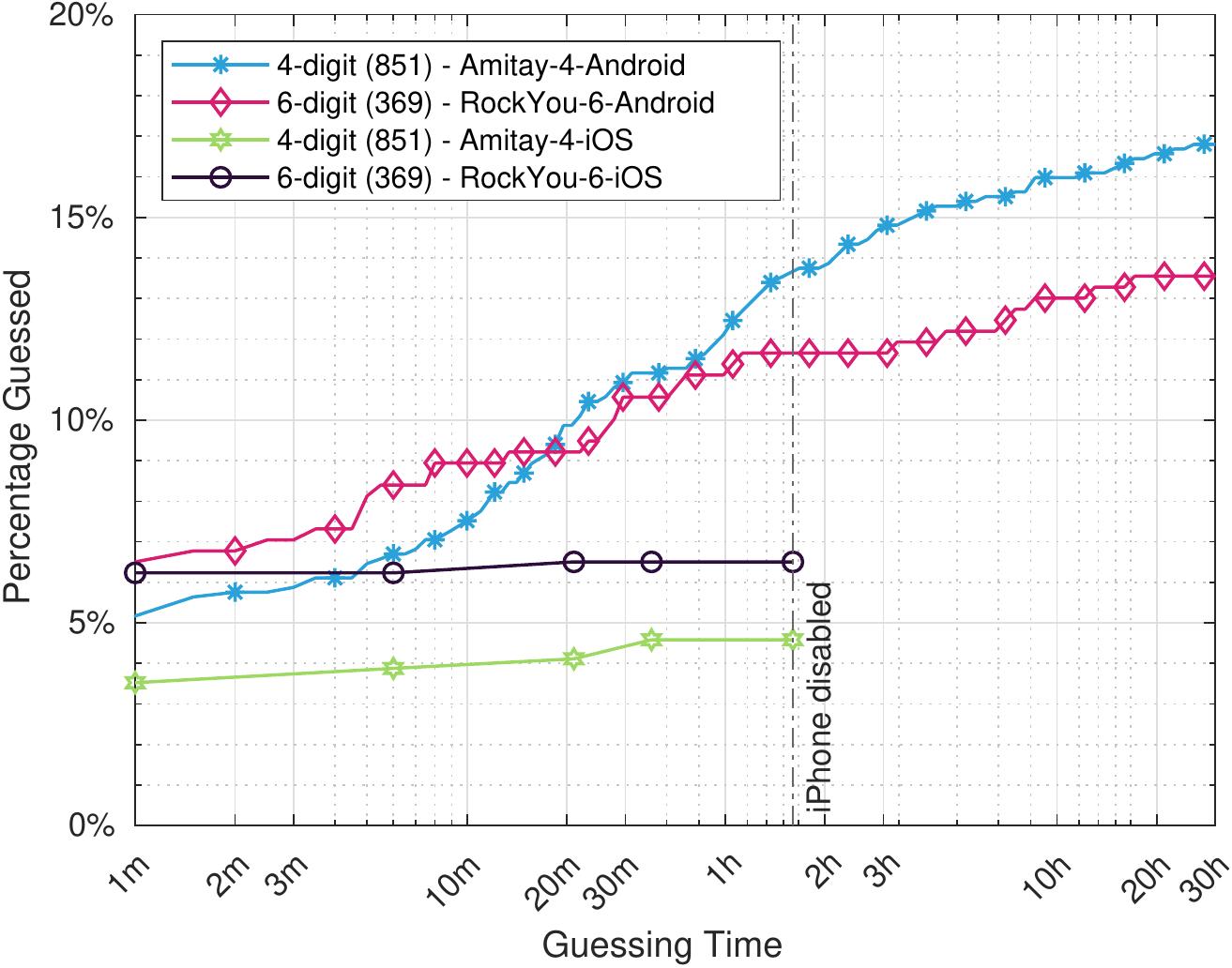}
    \label{fig:comparison-time}
}

\caption{Guessing performance of a \emph{throttled} attacker. The figure on the top is based on the number of guesses. The bottom figure is based on the required time and considers the different rate limits of Android and iOS (cf. Table~\ref{tab:rate-limiting}).}\label{fig:comparison}
\vspace{-1.5em}
\end{figure}

An overview of our guessing analysis can be found in Figure~\ref{fig:comparison}.
In the throttled scenario, depicted in Figure~\ref{fig:comparison-100}, we find attacking 4-digit~PINs with the \fourAmitay{} dataset~(\textcolor{darkgray}{$\triangle$}) is more effective than using \fourRockyou{}~(\textcolor{lightgray}{$\bigtriangledown$}).
We simulate the stronger attacker by utilizing the Amitay dataset in subsequent strength estimations of 4-digit PINs.

When comparing 4-~(\textcolor{darkgray}{$\triangle$}) and 6-digit PINs (\textcolor{darkgreen}{$\times$}), we see that guessing performance varies.
For 10~guesses (the maximum allowed under iOS), we find 4.6\,\% of the 4-digit and 6.5\,\% of the 6-digit PINs are guessed.
For 30~guesses (a less determined attacker on Android), 7.6\,\% of the 4-digit and 8.9\,\% of the 6-digit PINs are guessed and for 100~guesses (a reasonable upper bound on Android), 16.2\,\% of the 4-digit and 13.3\,\% of the 6-digit~PINs.

Somewhat counter-intuitive is the weaker security for 6-digit PINs for the first 40~guesses.  Upon investigation, the most-common 6-digit PINs are more narrowly distributed than their most-common 4-digit counterparts.
The most common 6-digit PINs consist of simple PINs, such as \texttt{123456} as defined in Table~\ref{tab:selection-strategies} in Appendix~\ref{app:selectiong-changing}, and repeating digits. In contrast, the most common 4-digit PINs consist of simple PINs, patterns, dates, and repeating digits. As a result, the most common 6-digit PINs may actually be easier to guess and less diverse than the most common 4-digit PINs.

There could be many explanations for this counter-intuitive finding. One explanation may be that users have more 4-digit PIN sequences to draw on in choosing a PIN, such as dates, but have fewer natural 6-digit analogs, and thus revert to less diverse, more easily guessed choices. Another explanation may be that users have a false sense of security that comes with 6-digit PINs as they are ``two digits more secure'' than 4-digit PINs.
Thus, users do not feel that they need more complexity in their 6-digit PIN choices.
Either way, future research is needed to better understand this phenomenon, which has also been observed by Aviv et al.~\cite{aviv-15-pattern-bigger} in the context of increasing the size (4x4 vs. 3x3) of Android graphical unlock patterns.

Finally, we compare guessing resistance with other mobile authentication schemes including Android's graphical unlock patterns drawn on a 3x3 grid ({\small\textcolor{orange}{$\square$}}) and alphanumeric passwords ({\footnotesize\textcolor{red}{$\bigstar$}}), along with a uniform distribution of 3-digit PINs (\textcolor{blue}{--}).
In theory, a 3x3 grid allows 389\,112 unique patterns, yet, the distribution of patterns is highly skewed~\cite{uellenbeck-13-pattern}.
When considering an attack throttled to 100~guesses, 35.5\,\% of the patterns will be guessed.  Against this attack, 4- and 6-digit PINs are twice as good.
Password-based authentication, on the other hand, is the most secure scheme.
After 100 guesses only 1.9\,\% of the passwords are recovered.

Figure~\ref{fig:comparison-time} shows the guessing time of an attacker due to rate limiting based on Table~\ref{tab:rate-limiting} for iOS and Android.
iOS has stricter rate limiting with a maximum of 10 guesses that can be completed in 1h 36m, at which point an attacker compromises 4.6\,\% of the 4-digit PINs and 6.5\,\% of the 6-digit PINs.
At the same time limit of roughly 1.5\,h, an attacker on Android is able to compromise 13.6\,\% of the 4-digit PINs and 11.7\,\% of the 6-digit PINs because of less restrictive rate limiting.

Especially on iOS, rate limiting becomes more aggressive after the initial guesses. For example, the first 6 guesses on iOS can be done within a minute, while the first 8 guesses already take 21 minutes. An attacker with only one minute on iOS is able to compromise 3.5\,\% of the 4-digit PINs and 6.2\,\% of the 6-digit PINs. But there are only marginal gains for 10 guesses which take 1h 36m on iOS with 4.6\,\% of the 4-digit PINs and 6.5\,\% of 6-digit PINs compromised.
Hence, after the first minute with 6 guesses on iOS, it does not greatly benefit the attacker to continue through the aggressive timeouts for 4 more guesses at 1h 36m.
In contrast, an attacker on Android would benefit more from continuing to guess beyond the initial large increases in rate limiting.
Of course, in a targeted attack setting, there may be additional information or other motivations for the attacker not modeled here.

To summarize, in line with previous work from Wang et al.~\cite{wang-17-pin}, we found no evidence that 6-digit PINs offer any security advantage over 4-digit PINs considering a throttled guessing attacker, which is the relevant threat model for mobile unlock authentication.
To support this claim, we performed $\chi^{2}$~tests ($\alpha=0.05$) for both the 4- and 6-digit PINs guessed within 10~[4.6\,\%, 6.5\,\%], 30~[7.6\,\%, 8.9\,\%], and 100~guesses~[16.2\,\%, 13.3\,\%]. Neither the test for 10 guesses showed a significant difference ($p = 0.16$) in PIN strength, nor the tests for 30 ($p = 0.44$) or 100~guesses ($p = 0.19$).

%% file: 05-2-selection.tex
\subsection{Selection Strategies}

In Step~(6) of our study, we asked participants about their ``strategy for choosing'' their PIN.
We analyzed the free-text responses to this question by building a codebook from a random sample of 200~PIN selection strategies using two coders. Inter-rater reliability between the coders measured by Cohen's kappa was $\kappa = 0.92$.
The 10~most popular strategies are shown in Appendix~\ref{app:selectiong-changing} in Table~\ref{tab:selection-strategies}.
We found no difference in the top~5 selection strategies between 4- and 6-digit PINs.

While the set of selection strategies is diverse, we found that many of the participants chose their PINs based on dates, especially birthdays and anniversaries.
Followed by that are PINs that were perceived memorable by participants who have selected something ``easy to remember.''
Also popular are patterns on the PIN pad and PINs that have some meaning to the participants like a partial ZIP code or a favorite number.

\begin{table*}
    \caption{Security metrics and creation times for PINs considering different datasets and treatments.}\label{tab:treatments-results}
  \centering
  \small
\resizebox{0.85\textwidth}{!}{\begin{tabular}{ll|cc|rrrrrrr|rrc}
\toprule
\multicolumn{2}{c|}{~} & \multicolumn{1}{c}{~}
& \multicolumn{1}{c|}{\textbf{Blocklist}}
& \multicolumn{2}{c}{\textbf{10 Guesses}}
& \multicolumn{2}{c}{\textbf{30 Guesses}}
& \multicolumn{2}{c}{\textbf{100 Guesses}}
& \multicolumn{1}{c|}{\textbf{Guess No.}}
& \multicolumn{1}{c}{\textbf{Creation}}
& \multicolumn{1}{c}{\textbf{Entry}}
& \multicolumn{1}{c}{\textbf{Number of}} \\
\multicolumn{1}{c}{~}
& \multicolumn{1}{l|}{\textbf{Name}}
& \multicolumn{1}{c}{\textbf{Participants}}
& \multicolumn{1}{c|}{\textbf{Hits}}
& \multicolumn{1}{c}{\textbf{No.}}
& \multicolumn{1}{c}{\textbf{\%}}
& \multicolumn{1}{c}{\textbf{No.}}
& \multicolumn{1}{c}{\textbf{\%}}
& \multicolumn{1}{c}{\textbf{No.}}
& \multicolumn{1}{c}{\textbf{\%}}
& \multicolumn{1}{c|}{\textbf{Median}}
& \multicolumn{1}{c}{\textbf{Time}}
& \multicolumn{1}{c}{\textbf{Time}}
& \multicolumn{1}{c}{\textbf{Attempts}} \\
\midrule
\multirow{2}{*}{\rotatebox{90}{\scriptsize{\textbf{Datasets}}}} & \fourFirst{}       & 851 & \hspace{1em}- & 39 & 5\,\% & 65 & 8\,\% & 138 & 16\,\%  & 1\,330 & -\hspace{0.55em} & -\hspace{0.55em} & -  \\[2.5pt]
 & Clicked-through-4    & ~\,19 & ~\,19 & 5  & 26\,\% & 6 & 32\,\% & 13  & 68\,\%  & 50  & -\hspace{0.55em} & -\hspace{0.55em} & - \\[2.5pt]
\midrule
& \fourControl{}       & 231 & \hspace{1em}- & 11 & 5\,\% & 19 & 8\,\% & 39  & 17\,\%  & 1\,185 & 7.9\,s & 1.48\,s & 1.01 \\[2pt]
\multirow{5}{*}{\rotatebox{90}{\scriptsize{\textbf{Treatments}}}} & \fourPlacebo{}       & 122 & 122 & 5  & 4\,\% & 11 & 9\,\%  & 19  & 16\,\%  & 2\,423 & 21.8\,s & 1.52\,s & 2.15 \\[2pt]
& \fourIosWith{}       & 124 & ~\,28 & 5  & 4\,\% & 8 & 6\,\%  & 18 & 15\,\%  & 1\,405 & 10.4\,s & 1.36\,s & 1.17 \\[2pt]
& \fourIosNo{}         & 126 & ~\,21 & 4 & 3\,\% & 10 & 8\,\% & 14 & 11\,\%  & 1\,747 & ~\,9.3\,s & 1.58\,s & 1.29  \\[2pt]
& \fourDataSmall{}     & 121 & \hspace{0.85em}5 & 4 & 3\,\% & 7 & 6\,\% & 18 & 15\,\%  & 1\,928 & ~\,8.8\,s & 1.47\,s & 1.11 \\[2pt]
& \fourDataLarge{}     & 127 & ~\,88 & 0 & 0\,\% & 0 & 0\,\% & 1 & 1\,\%  & 2\,871 & 25.4\,s & 1.55\,s & 2.98  \\[2pt]
\midrule
\multirow{2}{*}{\rotatebox{90}{\scriptsize{\textbf{Datasets}}}} & \sixFirst{}       & 369 & \hspace{1em}- & 24 & 7\,\% & 33 & 9\,\% & 49  & 13\,\%  & 39\,389  & -\hspace{0.55em} & -\hspace{0.55em} & -  \\[2.5pt]
& Clicked-through-6    & ~\,10 & ~\,10 & 9  &  90\,\% & 9 & 90\,\% &  9   & 90\,\%  & 1     & -\hspace{0.55em} & -\hspace{0.55em} & -  \\[2.5pt]
\midrule
\multirow{3}{*}{\rotatebox{90}{\scriptsize{\textbf{Treatments}}}} & \sixControl{}        & 127 & \hspace{1em}- & 7  & 6\,\% & 12 & 9\,\%  & 18  & 14\,\%  & 36\,822  & 11.5\,s & 2.52\,s & 1.01 \\[2pt]
& \sixPlacebo{}        & 117 & 117 & 3  & 3\,\% & 6 & 5\,\%  & 10  & 9\,\%   & 154\,521 & 28.5\,s & 2.98\,s & 2.17\\[2pt]
& \sixIosWith{}        & 125 & ~\,15 & 9  & 7\,\% & 9 & 7\,\%  & 13  & 10\,\%  & 40\,972  & 11.9\,s & 2.56\,s & 1.06 \\[2pt]
\bottomrule
\end{tabular}}
    \vspace{-.2in}
\end{table*}

%% file: 06-blacklists.tex
\section{Blocklists and PIN Selection}
\label{sec:blocklists}
\vspace{-.01in}
We now present results on
our 7 blocklist treatments: 5 treatments for 4-digit PINs and 2
treatments for 6-digit PINs as shown in Table~\ref{tab:treatments-results}.

\input{06-1-attacker-knowledge}
\input{06-2-blacklist-impact}
\input{06-3-blacklist-enforcement}
\input{06-4-changing}
\input{06-5-perception}
\input{06-6-sentiment} 

%% file: 06-1-attacker-knowledge.tex
\subsection{Attacker's Knowledge of Blocklists}\label{sec:attacker-knowledge}

As described in Section~\ref{sec:attacker-model}, we assume the attacker knows which blocking strategy is used by the system and can optimize
the guessing strategy by {\em not} guessing items on the blocklist.
Here, we consider how much
benefit this optimization provides.
Table~\ref{tab:security-delta} shows the net gains and losses for guessing PINs when
considering a blocklist-informed attacker.

Knowledge of the blocklist is unhelpful when considering the
placebo (\fourPlaceboAbbr{} and \sixPlaceboAbbr{}) and the click-through treatments
(\fourIosWithAbbr{} and \sixIosWithAbbr{}).
The blocklist is effectively of size one for the placebo
as the first choice of a participant is dynamically blocked.
Merely knowing that a PIN was blocked is of little help to the attacker.
As there is no clear gain (or harm), we model a blocklist-knowledgeable attacker for the placebo treatments
(see Table~\ref{tab:treatments-results}).

The case with a non-enforcing blocklist where users can click through the
warning message is more subtle. If the attacker is explicitly choosing not to
consider PINs on the blocklist, even though they may {\em actually} be selected
due to non-enforcement, the guessing strategy is harmed (negative in
Table~\ref{tab:security-delta}).
None of the tested modifications of this strategy, e.g. by incorporating the observed click-through rate, lead to an improvement.
As such, we consider an attacker that {\em does not} use the blocklist to change their guessing strategy for the click-through treatments (\fourIosWithAbbr{} and \sixIosWithAbbr{}).
In the remaining treatments (\fourIosNoAbbr{}, \fourDataSmallAbbr{}, \fourDataLargeAbbr{}), there are clear advantages when knowing the blocklist. 

%% file: 06-2-blacklist-impact.tex
\subsection{Blocklisting Impact on Security}
We now consider how the different blocklists perform in terms of improving
security.
The primary results are in Table~\ref{tab:treatments-results} where we report on the guessing performance against each treatment.
As described in Section~\ref{sec:attacker-model}, there are certain rate limits implemented on Android and iOS which is why we report on throttled attacks with 10, 30, and 100~guesses in
terms of the number and percentage of correctly guessed PINs (No. and \% columns).
In addition, we provide the attacker's performance in an unthrottled setting based on the median guess number.
The 4-digit attacker is informed by the \fourAmitayAbbr{}{}
dataset, while the 6-digit attacker employs the \sixRockyouAbbr{}{} dataset.
Both attackers guess in frequency order with knowledge of the blocklist where appropriate (see Section~\ref{sec:attacker-knowledge}).

We performed a multivariant $\chi^2$~test comparison ($\alpha=0.05$) for the PINs guessed within 10, 30, and 100~guesses across treatments. The test for 10 and 30~guesses did not show any significant difference ($p = 0.21$ and $p = 0.10$); the test for 100~guesses did ($p < 0.01$), as described below.

\begin{table}
	\vspace{-.2em}
    \caption{Attacker's gain from blocklist knowledge.}\label{tab:security-delta}
    \centering
\resizebox{1.0\columnwidth}{!}{\begin{tabular}{l|rrrrrrrc}
\toprule
\multicolumn{1}{c|}{~}
& \multicolumn{2}{c}{\textbf{10 Guesses}}
& \multicolumn{2}{c}{\textbf{30 Guesses}}
& \multicolumn{2}{c}{\textbf{100 Guesses}}
& \multicolumn{1}{c}{\textbf{Guess No.}}
& \multicolumn{1}{c}{\textbf{Knowledge}} \\
 \multicolumn{1}{l|}{\textbf{Treatment}}
 & \multicolumn{1}{c}{\textbf{No.}}
 & \multicolumn{1}{c}{\textbf{\%}}
 & \multicolumn{1}{c}{\textbf{No.}}
 & \multicolumn{1}{c}{\textbf{\%}}
 & \multicolumn{1}{c}{\textbf{No.}}
 & \multicolumn{1}{c}{\textbf{\%}}
 & \multicolumn{1}{c}{\textbf{Median}}
 & \multicolumn{1}{c}{\textbf{Beneficial}} \\
\midrule
\fourPlaceboAbbr{} & $\pm$0 & $\pm$0\,\% & $\pm$0 & $\pm$0\,\% & $\pm$0 & $\pm$0\,\% & $\pm$0 & \textbf{--} \\
\fourIosWithAbbr{}      &              -3 &              -2\,\% & -4 & -2\,\% &         -9 &          -8\,\% &           -303 & \xmark \\
\textbf{\fourIosNoAbbr{}}        &     \textbf{+3} &     \textbf{+2\,\%} &     \textbf{+7} &     \textbf{+6\,\%} & \textbf{+3} & \textbf{+2\,\%} &  \textbf{+245} & \cmark \\
\textbf{\fourDataSmallAbbr{}}    &     \textbf{+4} &     \textbf{+3\,\%} &     \textbf{+7} &     \textbf{+6\,\%} & \textbf{+5} & \textbf{+4\,\%} &   \textbf{+27} & \cmark \\
\textbf{\fourDataLargeAbbr{}}    & \textbf{$\pm$0} & \textbf{$\pm$0\,\%}& \textbf{$\pm$0} & \textbf{$\pm$0\,\%} & \textbf{+1} & \textbf{+1\,\%} & \textbf{+2740} & \cmark \\
\midrule
\sixPlaceboAbbr{}       &  $\pm$0 &  $\pm$0\,\% & $\pm$0 &  $\pm$0\,\% &  $\pm$0 &  $\pm$0\,\% &     $\pm$0 & \textbf{--} \\
\sixIosWithAbbr{}       &      -9 &      -7\,\% &      -5 &      -4\,\% &      -8 &      -6\,\% &      -7322 & \xmark \\
\bottomrule
\end{tabular}}
\vspace{-2em}
\end{table}

\paragraph{Smaller Blocklists} In the throttled setting with 100~guesses, there is little difference among \fourIosWith{} (15\,\%), \fourIosNo{} (11\,\%), \fourDataSmall{} (15\,\%), \fourPlacebo{} (16\,\%), compared to \fourControl{} (17\,\%) and \fourFirst{} (16\,\%).
Our post-hoc analyses (Bonferroni-corrected for multiple testing) results support this, as we found no significant difference between the smaller blocklists.
It is therefore hard to justify the combination of throttling and small blocklists, especially as blocklist warnings are associated with negative sentiments (see Section~\ref{sec:sentiment}).

In the unthrottled setting, though, we see some differences between the smaller and placebo blocklist cases. Notably, the smallest blocklist (\fourDataSmall{})
outperforms the $10\times$ larger iOS blocklist (\fourIosNo{}). We conjecture
this may be due to iOS' inclusion of PINs based on repetitions which
were chosen less often by our participants.
As a result, in an unthrottled setting, blocking can offer real benefits.
The median guess numbers for both 4- and 6-digit placebos suggest that just pushing users away from their first choice can improve security.
Unfortunately, direct use of a placebo blocklist is unlikely to be effective and is problematic in practice as users will quickly figure out the deception.

Finally, we reiterate that these improvements to the unthrottled attack setting appear to be only of academic interest: given the small key space, it is reasonable to assume that all possible combinations can be exhaustively tested within minutes~\cite{reed-18-graykey-malwarebytes}.

\paragraph{Large Blocklist}
We also consider a very large blocklist in the \fourDataLarge{} treatment
containing 2740~PINs, $10\times$ bigger than the 4-digit iOS blocklist and blocking
27.4\,\% of the key space. At this scale, we do see noticeable effects on
security in the throttled setting.  Even after 100~guesses, the attacker finds only 1\,\% of 4-digit PINs.
Our $\chi^{2}$~tests support this, for 100~guesses we found a significant difference ($p < 0.01$). For post-hoc analyses (Bonferroni-corrected) we found a significant difference between the large \fourDataLargeAbbr{} blocklist and \sixControlAbbr{} ($p < 0.01$) as well as all other 4-digit treatments: \fourControlAbbr{} ($p < 0.001$), \fourPlaceboAbbr{} ($p < 0.01$), \fourIosWithAbbr{} ($p < 0.01$), \fourIosNoAbbr{} ($p < 0.05$), and \fourDataSmallAbbr{} ($p < 0.01$).
This suggests that a larger blocklist can improve security in a throttled setting.

While similar positive security results are present for the unthrottled setting, we show in Section~\ref{sec:perception} that the larger blocklist also leads to a perceived lower usability, and thus it is important to balance the user experience with security gains.

\paragraph{Correctly Sizing a Blocklist} While there is a clear benefit to having a
large blocklist, it is important to consider the right size of a blocklist to
counteract negative usability and user experience issues. This leads to the
question: {\em Can a smaller blocklist provide similar benefits in the throttled setting and if so, what is an appropriately sized blocklist?}

Data from the \fourDataLarge{} treatment enables us to simulate how users would have responded to shorter blocklists. In our user study, we collected not only the final PIN accepted by the system, but also all $n-1$ intermediate (first-choice, second-choice, and so on) PINs rejected due to the blocklist. Consider a smaller blocklist that would have permitted choice $n-1$ to be the final PIN, rather than $n$. To simulate that smaller blocklist size, we use choice $n-1$.

\begin{figure}[t]
\centering
\includegraphics[width=0.87\columnwidth]{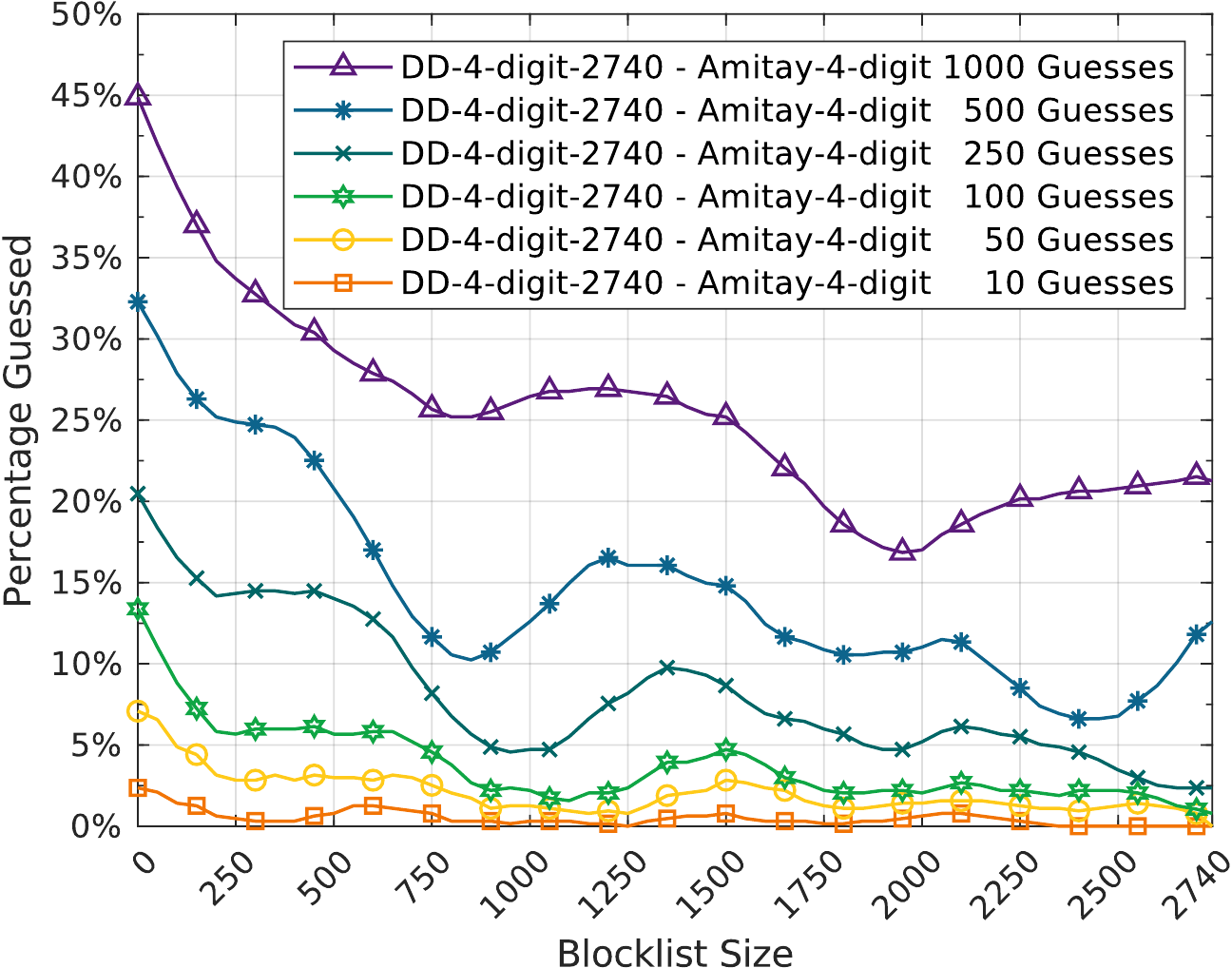}
\vspace{-.1in}
\caption{Blocklist size recommendation: For throttled attackers, limited to 100~guesses, a blocklist of $\sim\mkern-3mu 10$\,\% of the key space (\mbox{$\sim\mkern-3mu 1150$~PINs}) is ideal.}
\label{fig:recommendation}
\vspace{-.2in}
\end{figure}

The results of the simulation are shown in Figure~\ref{fig:recommendation}. We observe that there are several troughs and peaks in the curves.
We speculate that these relate to changes in user choices as they move
from their first choice PIN to their second choice PIN, and so on due to the
expanding blocklist restrictions. For example, entering the first
trough, the attacker is most disadvantaged when it is no longer possible to rely on guessing
only first choice PINs and second choice PINs need to be considered.
Eventually, the blocklist has restricted all first choice PINs, whereby the attacker can now take advantage of guessing popular second choices which results in a peak.
These cycles continue until the blocklist gets so large that few acceptable PINs remain,
and the attacker's advantage grows steadily by guessing the remaining PINs not on the blocklist.

Based on these cycles, we conclude that an appropriately-sized blocklist should
be based on one of the troughs where an attacker is most disadvantaged to maximize
the security gained in the throttled setting. As we are also concerned about
minimizing user discomfort and frustration (e.g, PIN creation time, see Table~\ref{tab:treatments-results}), the first trough appears the most ideal, which occurs at
about 10\,\% of the 4-digit PIN space throttled at 100 guesses.
We expect similar results for 6-digit blocklists as well, but we do not have sufficient examples of selected 6-digit PINs to perform the same analysis. Future research would be needed to confirm this. However, as 4-digit PINs are still exceedingly common, this result provides important insight to developers to improve user selection of 4-digit PINs.

\vspace{-.5em}

%% file: 06-3-blacklist-enforcement.tex
\input{fig/plots/perception/perception.tex}

\subsection{Enforcing the Blocklist}
In the 4-digit case, we compared the enforcing (\fourIosNo) with the non-enforcing (\fourIosWith{}) blocklist and found that enabling a click-through
option does not show significant security differences.
This suggests that using a click-through option does not reduce
security in the throttled attacker setting despite the fact that clicked-through
PINs are extremely weak (see row Clicked-through-4 in
Table~\ref{tab:treatments-results}).

These results seem to be driven by the fact that it is uncertain whether the user clicked through (see Table~\ref{tab:security-delta}).
In an enforcing setting, the attacker can leverage the blocklist but is equally challenged in guessing the remaining PINs.

We also investigated why participants chose to ignore and click through the warning messages.
From the 28 participants who saw a blocklist warning in the \fourIosWithAbbr{} treatment, we observed a click-through-rate (CTR) of 68\,\% (19 participants).
In the respective 6-digit treatment \sixIosWithAbbr{}, 10 out of 15, i.e., 67\,\%, ignored the warning.
This is twice the rate at which TLS warnings are ignored ($\sim30\,\%$)~\cite{stark-19-urlephant}.

Furthermore, we asked the 29~participants who pressed ``\emph{Use Anyway}'' about their motivations.
The 3 most observed answers are \emph{Memorability Issues:} ``Because this is the number I can remember,'' \emph{Incomplete Threat Models:} ``Many people don't tend to try the obvious PIN as they think it's too obvious so people won't use it,'' and \emph{Indifference}: ``I don't give [sic] about the warning. Security is overrated.''
These findings are similar to prior work where users do not follow external guidance for a number of reasons~\cite{lee-97-tough,wichmann-11-selfdetermintation,renaud-15-mental-models}.

In older versions of iOS, the blocklist warning message was ``\emph{Are You Sure You Want to Use This PIN? This PIN is commonly used and can be easily guessed.}'' with the safe option ``\emph{Choose New PIN}'' (bold) and the unsafe click-through option saying ``\emph{Use PIN}.''
We observed that Apple changed this wording with iOS 11 to what is depicted in Figure~\ref{fig:feedback:ct}.
Considering that TLS warning design research started with similarly high CTRs of around 70\,\%~\cite{akhawe-13-alice}, we hope that new designs can also improve blocklist warning CTRs~\cite{stark-19-urlephant}.

%% file: fig/plots/perception/perception.tex
\definecolor{secure}{HTML}{549C77} 
\definecolor{somewhat_secure}{HTML}{ABD4B4}
\definecolor{neither_secure_nor_insecure}{HTML}{A0A0A0}
\definecolor{somewhat_insecure}{HTML}{F9F5C3}
\definecolor{insecure}{HTML}{EBDF42} 

\definecolor{easy_to_remember}{HTML}{3598DB} 
\definecolor{somewhat_easy_to_remember}{HTML}{9ACCED}
\definecolor{neither_easy_nor_hard_to_remember}{HTML}{A0A0A0}
\definecolor{somewhat_hard_to_remember}{HTML}{EF959A}
\definecolor{difficult_to_remember}{HTML}{CE2029} 

\definecolor{easy_to_enter}{HTML}{950095} 
\definecolor{somewhat_easy_to_enter}{HTML}{E5B8FF}
\definecolor{neither_easy_nor_hard_to_enter}{HTML}{A0A0A0}
\definecolor{somewhat_hard_to_enter}{HTML}{FDC4A0}
\definecolor{difficult_to_enter}{HTML}{F26106} 

\begin{figure*}[!t]
\small
\renewcommand*{\arraystretch}{0.7}
	\centering
    \setlength{\tabcolsep}{0pt}
	\begin{tabular}{r c c c}
		\toprule
		\textbf{Treatment} & \textbf{Security} & \textbf{Memorability} & \textbf{Convenience} \\
		\midrule

\raisebox{0.1em}{\fourControlAbbr} & \scalebox{0.7}{
\begin{tikzpicture}
\begin{axis}[xbar stacked,bar width=1cm,ytick=\empty,y post scale=0.07,xmin=0,xmax=100,ymin=0,ymax=1,legend style={at={(0,-2.5)},
legend style={cells={align=left, anchor=center, fill}, nodes={inner sep=0.4ex,below=-2ex}},
column sep=0cm, draw=none, anchor=west, legend columns=5},xticklabel style={opacity=0,yshift=12pt}]
\addplot[fill=secure, xbar legend, mark=none] coordinates {(26.4069264069,1)};
\addplot[fill=somewhat_secure, xbar legend, mark=none] coordinates {(43.29004329,1)};
\addplot[fill=neither_secure_nor_insecure, xbar legend, mark=none] coordinates {(12.987012987,1)};
\addplot[fill=somewhat_insecure, xbar legend, mark=none] coordinates {(12.987012987,1)};
\addplot[fill=insecure, xbar legend, mark=none] coordinates {(4.329004329,1)};
\end{axis}
\end{tikzpicture}} & \scalebox{0.7}{
\begin{tikzpicture}
\begin{axis}[xbar stacked,bar width=1cm,ytick=\empty,y post scale=0.07,xmin=0,xmax=100,ymin=0,ymax=1,legend style={at={(0,-2.5)},
legend style={cells={align=left, anchor=center, fill}, nodes={inner sep=0.4ex,below=-2ex}},
column sep=0cm, draw=none, anchor=west, legend columns=5},xticklabel style={opacity=0,yshift=12pt}]
\addplot[fill=easy_to_remember, xbar legend, mark=none] coordinates {(69.2640692641,1)};
\addplot[fill=somewhat_easy_to_remember, xbar legend, mark=none] coordinates {(18.6147186147,1)};
\addplot[fill=neither_easy_nor_hard_to_remember, xbar legend, mark=none] coordinates {(7.79220779221,1)};
\addplot[fill=somewhat_hard_to_remember, xbar legend, mark=none] coordinates {(3.0303030303,1)};
\addplot[fill=difficult_to_remember, xbar legend, mark=none] coordinates {(1.2987012987,1)};
\end{axis}
\end{tikzpicture}} & \scalebox{0.7}{
\begin{tikzpicture}
\begin{axis}[xbar stacked,bar width=1cm,ytick=\empty,y post scale=0.07,xmin=0,xmax=100,ymin=0,ymax=1,legend style={at={(0,-2.5)},
legend style={cells={align=left, anchor=center, fill}, nodes={inner sep=0.4ex,below=-2ex}},
column sep=0cm, draw=none, anchor=west, legend columns=5},xticklabel style={opacity=0,yshift=12pt}]
\addplot[fill=easy_to_enter, xbar legend, mark=none] coordinates {(53.6796536797,1)};
\addplot[fill=somewhat_easy_to_enter, xbar legend, mark=none] coordinates {(21.2121212121,1)};
\addplot[fill=neither_easy_nor_hard_to_enter, xbar legend, mark=none] coordinates {(19.4805194805,1)};
\addplot[fill=somewhat_hard_to_enter, xbar legend, mark=none] coordinates {(4.329004329,1)};
\addplot[fill=difficult_to_enter, xbar legend, mark=none] coordinates {(1.2987012987,1)};
\end{axis}
\end{tikzpicture}}\\

\raisebox{0.1em}{\fourPlaceboAbbr} & \scalebox{0.7}{
\begin{tikzpicture}
\begin{axis}[xbar stacked,bar width=1cm,ytick=\empty,y post scale=0.07,xmin=0,xmax=100,ymin=0,ymax=1,legend style={at={(0,-2.5)},
legend style={cells={align=left, anchor=center, fill}, nodes={inner sep=0.4ex,below=-2ex}},
column sep=0cm, draw=none, anchor=west, legend columns=5},xticklabel style={opacity=0,yshift=12pt}]
\addplot[fill=secure, xbar legend, mark=none] coordinates {(38.5245901639,1)};
\addplot[fill=somewhat_secure, xbar legend, mark=none] coordinates {(42.6229508197,1)};
\addplot[fill=neither_secure_nor_insecure, xbar legend, mark=none] coordinates {(9.83606557377,1)};
\addplot[fill=somewhat_insecure, xbar legend, mark=none] coordinates {(6.55737704918,1)};
\addplot[fill=insecure, xbar legend, mark=none] coordinates {(2.45901639344,1)};
\end{axis}
\end{tikzpicture}} & \scalebox{0.7}{
\begin{tikzpicture}
\begin{axis}[xbar stacked,bar width=1cm,ytick=\empty,y post scale=0.07,xmin=0,xmax=100,ymin=0,ymax=1,legend style={at={(0,-2.5)},
legend style={cells={align=left, anchor=center, fill}, nodes={inner sep=0.4ex,below=-2ex}},
column sep=0cm, draw=none, anchor=west, legend columns=5},xticklabel style={opacity=0,yshift=12pt}]
\addplot[fill=easy_to_remember, xbar legend, mark=none] coordinates {(43.4426229508,1)};
\addplot[fill=somewhat_easy_to_remember, xbar legend, mark=none] coordinates {(36.0655737705,1)};
\addplot[fill=neither_easy_nor_hard_to_remember, xbar legend, mark=none] coordinates {(9.83606557377,1)};
\addplot[fill=somewhat_hard_to_remember, xbar legend, mark=none] coordinates {(6.55737704918,1)};
\addplot[fill=difficult_to_remember, xbar legend, mark=none] coordinates {(4.09836065574,1)};
\end{axis}
\end{tikzpicture}} & \scalebox{0.7}{
\begin{tikzpicture}
\begin{axis}[xbar stacked,bar width=1cm,ytick=\empty,y post scale=0.07,xmin=0,xmax=100,ymin=0,ymax=1,legend style={at={(0,-2.5)},
legend style={cells={align=left, anchor=center, fill}, nodes={inner sep=0.4ex,below=-2ex}},
column sep=0cm, draw=none, anchor=west, legend columns=5},xticklabel style={opacity=0,yshift=12pt}]
\addplot[fill=easy_to_enter, xbar legend, mark=none] coordinates {(48.3606557377,1)};
\addplot[fill=somewhat_easy_to_enter, xbar legend, mark=none] coordinates {(25.4098360656,1)};
\addplot[fill=neither_easy_nor_hard_to_enter, xbar legend, mark=none] coordinates {(14.7540983607,1)};
\addplot[fill=somewhat_hard_to_enter, xbar legend, mark=none] coordinates {(9.01639344262,1)};
\addplot[fill=difficult_to_enter, xbar legend, mark=none] coordinates {(2.45901639344,1)};
\end{axis}
\end{tikzpicture}}\\

\raisebox{0.1em}{\fourIosNoAbbr} & \scalebox{0.7}{
\begin{tikzpicture}
\begin{axis}[xbar stacked,bar width=1cm,ytick=\empty,y post scale=0.07,xmin=0,xmax=100,ymin=0,ymax=1,legend style={at={(0,-2.5)},
legend style={cells={align=left, anchor=center, fill}, nodes={inner sep=0.4ex,below=-2ex}},
column sep=0cm, draw=none, anchor=west, legend columns=5},xticklabel style={opacity=0,yshift=12pt}]
\addplot[fill=secure, xbar legend, mark=none] coordinates {(31.746031746,1)};
\addplot[fill=somewhat_secure, xbar legend, mark=none] coordinates {(50.0,1)};
\addplot[fill=neither_secure_nor_insecure, xbar legend, mark=none] coordinates {(11.1111111111,1)};
\addplot[fill=somewhat_insecure, xbar legend, mark=none] coordinates {(6.34920634921,1)};
\addplot[fill=insecure, xbar legend, mark=none] coordinates {(0.793650793651,1)};
\end{axis}
\end{tikzpicture}} & \scalebox{0.7}{
\begin{tikzpicture}
\begin{axis}[xbar stacked,bar width=1cm,ytick=\empty,y post scale=0.07,xmin=0,xmax=100,ymin=0,ymax=1,legend style={at={(0,-2.5)},
legend style={cells={align=left, anchor=center, fill}, nodes={inner sep=0.4ex,below=-2ex}},
column sep=0cm, draw=none, anchor=west, legend columns=5},xticklabel style={opacity=0,yshift=12pt}]
\addplot[fill=easy_to_remember, xbar legend, mark=none] coordinates {(66.6666666667,1)};
\addplot[fill=somewhat_easy_to_remember, xbar legend, mark=none] coordinates {(20.6349206349,1)};
\addplot[fill=neither_easy_nor_hard_to_remember, xbar legend, mark=none] coordinates {(5.55555555556,1)};
\addplot[fill=somewhat_hard_to_remember, xbar legend, mark=none] coordinates {(6.34920634921,1)};
\addplot[fill=difficult_to_remember, xbar legend, mark=none] coordinates {(0.793650793651,1)};
\end{axis}
\end{tikzpicture}} & \scalebox{0.7}{
\begin{tikzpicture}
\begin{axis}[xbar stacked,bar width=1cm,ytick=\empty,y post scale=0.07,xmin=0,xmax=100,ymin=0,ymax=1,legend style={at={(0,-2.5)},
legend style={cells={align=left, anchor=center, fill}, nodes={inner sep=0.4ex,below=-2ex}},
column sep=0cm, draw=none, anchor=west, legend columns=5},xticklabel style={opacity=0,yshift=12pt}]
\addplot[fill=easy_to_enter, xbar legend, mark=none] coordinates {(57.1428571429,1)};
\addplot[fill=somewhat_easy_to_enter, xbar legend, mark=none] coordinates {(20.6349206349,1)};
\addplot[fill=neither_easy_nor_hard_to_enter, xbar legend, mark=none] coordinates {(14.2857142857,1)};
\addplot[fill=somewhat_hard_to_enter, xbar legend, mark=none] coordinates {(7.14285714286,1)};
\addplot[fill=difficult_to_enter, xbar legend, mark=none] coordinates {(0.793650793651,1)};
\end{axis}
\end{tikzpicture}}\\

\raisebox{0.1em}{\fourIosWithAbbr} & \scalebox{0.7}{
\begin{tikzpicture}
\begin{axis}[xbar stacked,bar width=1cm,ytick=\empty,y post scale=0.07,xmin=0,xmax=100,ymin=0,ymax=1,legend style={at={(0,-2.5)},
legend style={cells={align=left, anchor=center, fill}, nodes={inner sep=0.4ex,below=-2ex}},
column sep=0cm, draw=none, anchor=west, legend columns=5},xticklabel style={opacity=0,yshift=12pt}]
\addplot[fill=secure, xbar legend, mark=none] coordinates {(33.8709677419,1)};
\addplot[fill=somewhat_secure, xbar legend, mark=none] coordinates {(41.935483871,1)};
\addplot[fill=neither_secure_nor_insecure, xbar legend, mark=none] coordinates {(14.5161290323,1)};
\addplot[fill=somewhat_insecure, xbar legend, mark=none] coordinates {(8.87096774194,1)};
\addplot[fill=insecure, xbar legend, mark=none] coordinates {(0.806451612903,1)};
\end{axis}
\end{tikzpicture}} & \scalebox{0.7}{
\begin{tikzpicture}
\begin{axis}[xbar stacked,bar width=1cm,ytick=\empty,y post scale=0.07,xmin=0,xmax=100,ymin=0,ymax=1,legend style={at={(0,-2.5)},
legend style={cells={align=left, anchor=center, fill}, nodes={inner sep=0.4ex,below=-2ex}},
column sep=0cm, draw=none, anchor=west, legend columns=5},xticklabel style={opacity=0,yshift=12pt}]
\addplot[fill=easy_to_remember, xbar legend, mark=none] coordinates {(71.7741935484,1)};
\addplot[fill=somewhat_easy_to_remember, xbar legend, mark=none] coordinates {(16.935483871,1)};
\addplot[fill=neither_easy_nor_hard_to_remember, xbar legend, mark=none] coordinates {(6.45161290323,1)};
\addplot[fill=somewhat_hard_to_remember, xbar legend, mark=none] coordinates {(4.03225806452,1)};
\addplot[fill=difficult_to_remember, xbar legend, mark=none] coordinates {(0.806451612903,1)};
\end{axis}
\end{tikzpicture}} & \scalebox{0.7}{
\begin{tikzpicture}
\begin{axis}[xbar stacked,bar width=1cm,ytick=\empty,y post scale=0.07,xmin=0,xmax=100,ymin=0,ymax=1,legend style={at={(0,-2.5)},
legend style={cells={align=left, anchor=center, fill}, nodes={inner sep=0.4ex,below=-2ex}},
column sep=0cm, draw=none, anchor=west, legend columns=5},xticklabel style={opacity=0,yshift=12pt}]
\addplot[fill=easy_to_enter, xbar legend, mark=none] coordinates {(57.2580645161,1)};
\addplot[fill=somewhat_easy_to_enter, xbar legend, mark=none] coordinates {(18.5483870968,1)};
\addplot[fill=neither_easy_nor_hard_to_enter, xbar legend, mark=none] coordinates {(13.7096774194,1)};
\addplot[fill=somewhat_hard_to_enter, xbar legend, mark=none] coordinates {(8.06451612903,1)};
\addplot[fill=difficult_to_enter, xbar legend, mark=none] coordinates {(2.41935483871,1)};
\end{axis}
\end{tikzpicture}}\\

\raisebox{0.1em}{\fourDataSmallAbbr} & \scalebox{0.7}{
\begin{tikzpicture}
\begin{axis}[xbar stacked,bar width=1cm,ytick=\empty,y post scale=0.07,xmin=0,xmax=100,ymin=0,ymax=1,legend style={at={(0,-2.5)},
legend style={cells={align=left, anchor=center, fill}, nodes={inner sep=0.4ex,below=-2ex}},
column sep=0cm, draw=none, anchor=west, legend columns=5},xticklabel style={opacity=0,yshift=12pt}]
\addplot[fill=secure, xbar legend, mark=none] coordinates {(28.0991735537,1)};
\addplot[fill=somewhat_secure, xbar legend, mark=none] coordinates {(49.5867768595,1)};
\addplot[fill=neither_secure_nor_insecure, xbar legend, mark=none] coordinates {(6.61157024793,1)};
\addplot[fill=somewhat_insecure, xbar legend, mark=none] coordinates {(12.3966942149,1)};
\addplot[fill=insecure, xbar legend, mark=none] coordinates {(3.30578512397,1)};
\end{axis}
\end{tikzpicture}} & \scalebox{0.7}{
\begin{tikzpicture}
\begin{axis}[xbar stacked,bar width=1cm,ytick=\empty,y post scale=0.07,xmin=0,xmax=100,ymin=0,ymax=1,legend style={at={(0,-2.5)},
legend style={cells={align=left, anchor=center, fill}, nodes={inner sep=0.4ex,below=-2ex}},
column sep=0cm, draw=none, anchor=west, legend columns=5},xticklabel style={opacity=0,yshift=12pt}]
\addplot[fill=easy_to_remember, xbar legend, mark=none] coordinates {(64.4628099174,1)};
\addplot[fill=somewhat_easy_to_remember, xbar legend, mark=none] coordinates {(23.1404958678,1)};
\addplot[fill=neither_easy_nor_hard_to_remember, xbar legend, mark=none] coordinates {(8.26446280992,1)};
\addplot[fill=somewhat_hard_to_remember, xbar legend, mark=none] coordinates {(3.30578512397,1)};
\addplot[fill=difficult_to_remember, xbar legend, mark=none] coordinates {(0.826446280992,1)};
\end{axis}
\end{tikzpicture}} & \scalebox{0.7}{
\begin{tikzpicture}
\begin{axis}[xbar stacked,bar width=1cm,ytick=\empty,y post scale=0.07,xmin=0,xmax=100,ymin=0,ymax=1,legend style={at={(0,-2.5)},
legend style={cells={align=left, anchor=center, fill}, nodes={inner sep=0.4ex,below=-2ex}},
column sep=0cm, draw=none, anchor=west, legend columns=5},xticklabel style={opacity=0,yshift=12pt}]
\addplot[fill=easy_to_enter, xbar legend, mark=none] coordinates {(53.7190082645,1)};
\addplot[fill=somewhat_easy_to_enter, xbar legend, mark=none] coordinates {(27.2727272727,1)};
\addplot[fill=neither_easy_nor_hard_to_enter, xbar legend, mark=none] coordinates {(11.5702479339,1)};
\addplot[fill=somewhat_hard_to_enter, xbar legend, mark=none] coordinates {(6.61157024793,1)};
\addplot[fill=difficult_to_enter, xbar legend, mark=none] coordinates {(0.826446280992,1)};
\end{axis}
\end{tikzpicture}}\\

\raisebox{0.1em}{\fourDataLargeAbbr} & \scalebox{0.7}{
\begin{tikzpicture}
\begin{axis}[xbar stacked,bar width=1cm,ytick=\empty,y post scale=0.07,xmin=0,xmax=100,ymin=0,ymax=1,legend style={at={(0,-2.5)},
legend style={cells={align=left, anchor=center, fill}, nodes={inner sep=0.4ex,below=-2ex}},
column sep=0cm, draw=none, anchor=west, legend columns=5},xticklabel style={opacity=0,yshift=12pt}]
\addplot[fill=secure, xbar legend, mark=none] coordinates {(43.3070866142,1)};
\addplot[fill=somewhat_secure, xbar legend, mark=none] coordinates {(43.3070866142,1)};
\addplot[fill=neither_secure_nor_insecure, xbar legend, mark=none] coordinates {(10.2362204724,1)};
\addplot[fill=somewhat_insecure, xbar legend, mark=none] coordinates {(2.36220472441,1)};
\addplot[fill=insecure, xbar legend, mark=none] coordinates {(0.787401574803,1)};
\end{axis}
\end{tikzpicture}} & \scalebox{0.7}{
\begin{tikzpicture}
\begin{axis}[xbar stacked,bar width=1cm,ytick=\empty,y post scale=0.07,xmin=0,xmax=100,ymin=0,ymax=1,legend style={at={(0,-2.5)},
legend style={cells={align=left, anchor=center, fill}, nodes={inner sep=0.4ex,below=-2ex}},
column sep=0cm, draw=none, anchor=west, legend columns=5},xticklabel style={opacity=0,yshift=12pt}]
\addplot[fill=easy_to_remember, xbar legend, mark=none] coordinates {(44.094488189,1)};
\addplot[fill=somewhat_easy_to_remember, xbar legend, mark=none] coordinates {(22.0472440945,1)};
\addplot[fill=neither_easy_nor_hard_to_remember, xbar legend, mark=none] coordinates {(11.0236220472,1)};
\addplot[fill=somewhat_hard_to_remember, xbar legend, mark=none] coordinates {(15.7480314961,1)};
\addplot[fill=difficult_to_remember, xbar legend, mark=none] coordinates {(7.08661417323,1)};
\end{axis}
\end{tikzpicture}} & \scalebox{0.7}{
\begin{tikzpicture}
\begin{axis}[xbar stacked,bar width=1cm,ytick=\empty,y post scale=0.07,xmin=0,xmax=100,ymin=0,ymax=1,legend style={at={(0,-2.5)},
legend style={cells={align=left, anchor=center, fill}, nodes={inner sep=0.4ex,below=-2ex}},
column sep=0cm, draw=none, anchor=west, legend columns=5},xticklabel style={opacity=0,yshift=12pt}]
\addplot[fill=easy_to_enter, xbar legend, mark=none] coordinates {(39.3700787402,1)};
\addplot[fill=somewhat_easy_to_enter, xbar legend, mark=none] coordinates {(26.7716535433,1)};
\addplot[fill=neither_easy_nor_hard_to_enter, xbar legend, mark=none] coordinates {(17.3228346457,1)};
\addplot[fill=somewhat_hard_to_enter, xbar legend, mark=none] coordinates {(12.5984251969,1)};
\addplot[fill=difficult_to_enter, xbar legend, mark=none] coordinates {(3.93700787402,1)};
\end{axis}
\end{tikzpicture}}\\

\raisebox{0.1em}{\sixControlAbbr} & \scalebox{0.7}{
\begin{tikzpicture}
\begin{axis}[xbar stacked,bar width=1cm,ytick=\empty,y post scale=0.07,xmin=0,xmax=100,ymin=0,ymax=1,legend style={at={(0,-2.5)},
legend style={cells={align=left, anchor=center, fill}, nodes={inner sep=0.4ex,below=-2ex}},
column sep=0cm, draw=none, anchor=west, legend columns=5},xticklabel style={opacity=0,yshift=12pt}]
\addplot[fill=secure, xbar legend, mark=none] coordinates {(31.4960629921,1)};
\addplot[fill=somewhat_secure, xbar legend, mark=none] coordinates {(44.094488189,1)};
\addplot[fill=neither_secure_nor_insecure, xbar legend, mark=none] coordinates {(10.2362204724,1)};
\addplot[fill=somewhat_insecure, xbar legend, mark=none] coordinates {(11.0236220472,1)};
\addplot[fill=insecure, xbar legend, mark=none] coordinates {(3.14960629921,1)};
\end{axis}
\end{tikzpicture}} & \scalebox{0.7}{
\begin{tikzpicture}
\begin{axis}[xbar stacked,bar width=1cm,ytick=\empty,y post scale=0.07,xmin=0,xmax=100,ymin=0,ymax=1,legend style={at={(0,-2.5)},
legend style={cells={align=left, anchor=center, fill}, nodes={inner sep=0.4ex,below=-2ex}},
column sep=0cm, draw=none, anchor=west, legend columns=5},xticklabel style={opacity=0,yshift=12pt}]
\addplot[fill=easy_to_remember, xbar legend, mark=none] coordinates {(68.5039370079,1)};
\addplot[fill=somewhat_easy_to_remember, xbar legend, mark=none] coordinates {(22.8346456693,1)};
\addplot[fill=neither_easy_nor_hard_to_remember, xbar legend, mark=none] coordinates {(4.72440944882,1)};
\addplot[fill=somewhat_hard_to_remember, xbar legend, mark=none] coordinates {(3.14960629921,1)};
\addplot[fill=difficult_to_remember, xbar legend, mark=none] coordinates {(0.787401574803,1)};
\end{axis}
\end{tikzpicture}} & \scalebox{0.7}{
\begin{tikzpicture}
\begin{axis}[xbar stacked,bar width=1cm,ytick=\empty,y post scale=0.07,xmin=0,xmax=100,ymin=0,ymax=1,legend style={at={(0,-2.5)},
legend style={cells={align=left, anchor=center, fill}, nodes={inner sep=0.4ex,below=-2ex}},
column sep=0cm, draw=none, anchor=west, legend columns=5},xticklabel style={opacity=0,yshift=12pt}]
\addplot[fill=easy_to_enter, xbar legend, mark=none] coordinates {(60.6299212598,1)};
\addplot[fill=somewhat_easy_to_enter, xbar legend, mark=none] coordinates {(23.6220472441,1)};
\addplot[fill=neither_easy_nor_hard_to_enter, xbar legend, mark=none] coordinates {(9.44881889764,1)};
\addplot[fill=somewhat_hard_to_enter, xbar legend, mark=none] coordinates {(3.93700787402,1)};
\addplot[fill=difficult_to_enter, xbar legend, mark=none] coordinates {(2.36220472441,1)};
\end{axis}
\end{tikzpicture}}\\

\raisebox{0.1em}{\sixPlaceboAbbr} & \scalebox{0.7}{
\begin{tikzpicture}
\begin{axis}[xbar stacked,bar width=1cm,ytick=\empty,y post scale=0.07,xmin=0,xmax=100,ymin=0,ymax=1,legend style={at={(0,-2.5)},
legend style={cells={align=left, anchor=center, fill}, nodes={inner sep=0.4ex,below=-2ex}},
column sep=0cm, draw=none, anchor=west, legend columns=5},xticklabel style={opacity=0,yshift=12pt}]
\addplot[fill=secure, xbar legend, mark=none] coordinates {(34.188034188,1)};
\addplot[fill=somewhat_secure, xbar legend, mark=none] coordinates {(53.8461538462,1)};
\addplot[fill=neither_secure_nor_insecure, xbar legend, mark=none] coordinates {(6.83760683761,1)};
\addplot[fill=somewhat_insecure, xbar legend, mark=none] coordinates {(5.12820512821,1)};
\addplot[fill=insecure, xbar legend, mark=none] coordinates {(0.0,1)};
\end{axis}
\end{tikzpicture}} & \scalebox{0.7}{
\begin{tikzpicture}
\begin{axis}[xbar stacked,bar width=1cm,ytick=\empty,y post scale=0.07,xmin=0,xmax=100,ymin=0,ymax=1,legend style={at={(0,-2.5)},
legend style={cells={align=left, anchor=center, fill}, nodes={inner sep=0.4ex,below=-2ex}},
column sep=0cm, draw=none, anchor=west, legend columns=5},xticklabel style={opacity=0,yshift=12pt}]
\addplot[fill=easy_to_remember, xbar legend, mark=none] coordinates {(44.4444444444,1)};
\addplot[fill=somewhat_easy_to_remember, xbar legend, mark=none] coordinates {(30.7692307692,1)};
\addplot[fill=neither_easy_nor_hard_to_remember, xbar legend, mark=none] coordinates {(8.54700854701,1)};
\addplot[fill=somewhat_hard_to_remember, xbar legend, mark=none] coordinates {(15.3846153846,1)};
\addplot[fill=difficult_to_remember, xbar legend, mark=none] coordinates {(0.854700854701,1)};
\end{axis}
\end{tikzpicture}} & \scalebox{0.7}{
\begin{tikzpicture}
\begin{axis}[xbar stacked,bar width=1cm,ytick=\empty,y post scale=0.07,xmin=0,xmax=100,ymin=0,ymax=1,legend style={at={(0,-2.5)},
legend style={cells={align=left, anchor=center, fill}, nodes={inner sep=0.4ex,below=-2ex}},
column sep=0cm, draw=none, anchor=west, legend columns=5},xticklabel style={opacity=0,yshift=12pt}]
\addplot[fill=easy_to_enter, xbar legend, mark=none] coordinates {(36.7521367521,1)};
\addplot[fill=somewhat_easy_to_enter, xbar legend, mark=none] coordinates {(33.3333333333,1)};
\addplot[fill=neither_easy_nor_hard_to_enter, xbar legend, mark=none] coordinates {(17.094017094,1)};
\addplot[fill=somewhat_hard_to_enter, xbar legend, mark=none] coordinates {(11.1111111111,1)};
\addplot[fill=difficult_to_enter, xbar legend, mark=none] coordinates {(1.7094017094,1)};
\end{axis}
\end{tikzpicture}}\\

\raisebox{3.3em}{\sixIosWithAbbr} & \scalebox{0.7}{
\begin{tikzpicture}
\begin{axis}[xbar stacked,bar width=1cm,ytick=\empty,y post scale=0.07,xmin=0,xmax=100,ymin=0,ymax=1,legend style={at={(0,-2.5)},
legend style={cells={align=left, anchor=center, fill}, nodes={inner sep=0.4ex,below=-2ex}},
column sep=0cm, draw=none, anchor=west, legend columns=5}]
\addplot[fill=secure, xbar legend, mark=none] coordinates {(26.4,1)};
\addplot[fill=somewhat_secure, xbar legend, mark=none] coordinates {(44.8,1)};
\addplot[fill=neither_secure_nor_insecure, xbar legend, mark=none] coordinates {(10.4,1)};
\addplot[fill=somewhat_insecure, xbar legend, mark=none] coordinates {(12.8,1)};
\addplot[fill=insecure, xbar legend, mark=none] coordinates {(5.6,1)};
\addlegendentry{Secure};
\addlegendentry{Somew.\\secure};
\addlegendentry{Neither};
\addlegendentry{Somew.\\insecure};
\addlegendentry{Insecure};
\end{axis}
\end{tikzpicture}} & \scalebox{0.7}{
\begin{tikzpicture}
\begin{axis}[xbar stacked,bar width=1cm,ytick=\empty,y post scale=0.07,xmin=0,xmax=100,ymin=0,ymax=1,legend style={at={(0,-2.5)},
legend style={cells={align=left, anchor=center, fill}, nodes={inner sep=0.4ex,below=-2ex}},
column sep=0cm, draw=none, anchor=west, legend columns=5}]
\addplot[fill=easy_to_remember, xbar legend, mark=none] coordinates {(60.8,1)};
\addplot[fill=somewhat_easy_to_remember, xbar legend, mark=none] coordinates {(29.6,1)};
\addplot[fill=neither_easy_nor_hard_to_remember, xbar legend, mark=none] coordinates {(6.4,1)};
\addplot[fill=somewhat_hard_to_remember, xbar legend, mark=none] coordinates {(2.4,1)};
\addplot[fill=difficult_to_remember, xbar legend, mark=none] coordinates {(0.8,1)};
\addlegendentry{Easy\\to rmb.};
\addlegendentry{Somew.\\easy};
\addlegendentry{Neither};
\addlegendentry{Somew.\\hard};
\addlegendentry{Difficult\\to rmb.};
\end{axis}
\end{tikzpicture}} & \scalebox{0.7}{
\begin{tikzpicture}
\begin{axis}[xbar stacked,bar width=1cm,ytick=\empty,y post scale=0.07,xmin=0,xmax=100,ymin=0,ymax=1,legend style={at={(0,-2.5)},
legend style={cells={align=left, anchor=center, fill}, nodes={inner sep=0.4ex,below=-2ex}},
column sep=0cm, draw=none, anchor=west, legend columns=5}]
\addplot[fill=easy_to_enter, xbar legend, mark=none] coordinates {(45.6,1)};
\addplot[fill=somewhat_easy_to_enter, xbar legend, mark=none] coordinates {(27.2,1)};
\addplot[fill=neither_easy_nor_hard_to_enter, xbar legend, mark=none] coordinates {(15.2,1)};
\addplot[fill=somewhat_hard_to_enter, xbar legend, mark=none] coordinates {(12.0,1)};
\addplot[fill=difficult_to_enter, xbar legend, mark=none] coordinates {(0.0,1)};
\addlegendentry{Easy to\\enter};
\addlegendentry{Somew.\\easy};
\addlegendentry{Neither};
\addlegendentry{Somew.\\hard};
\addlegendentry{Difficult\\to enter};
\end{axis}
\end{tikzpicture}}\\

		\bottomrule
	\end{tabular}
	\caption{Participants' perception of their PIN's security (\emph{Secure -- Insecure}), memorability (\emph{Easy to remember -- Difficult to remember}), and convenience (\emph{Easy to enter -- Difficult to enter}).}
	\label{fig:perception-sec-mem-con}
        \vspace{-.2in}
\end{figure*}
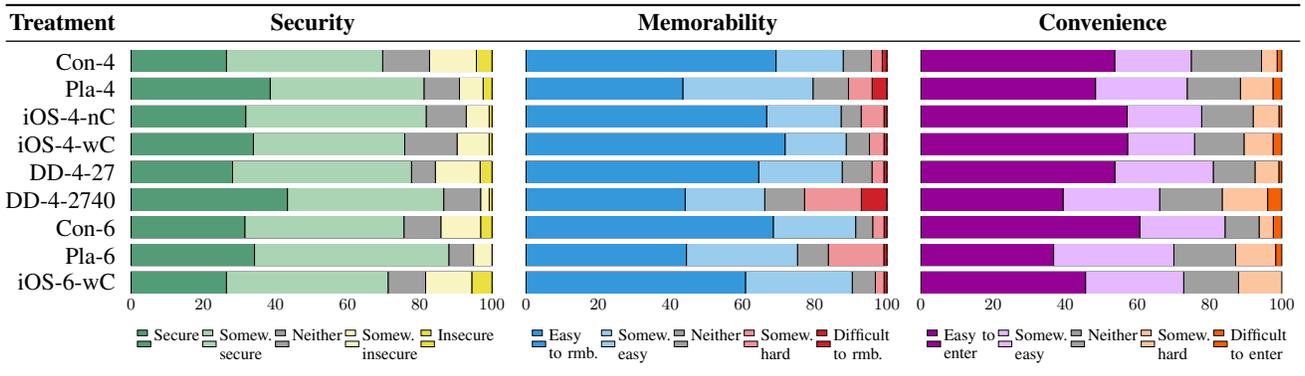

%% file: 06-4-changing.tex
\subsection{PIN Changing Strategies}\label{sec:pin-changing}
In our study, we asked 367 participants who faced a blocklist how their  creation strategy changed in response to the warning.
We sampled 126 responses (10\,\% of our total number of participants) and grouped them into three categories: participants who continued using the ``Same'' strategy, participants who made ``Minor'' changes to the strategy, and participants who came up with a completely ``New'' strategy.
Examples for those cases can be found in the Appendix~\ref{app:selectiong-changing} in Table~\ref{tab:changing-strategies-codebook}.
Two coders independently coded the data.
Inter-rater reliability between the coders measured by Cohen's kappa was $\kappa = 0.96$.
The results are shown in Table~\ref{tab:changing-strategies}.

About 50\,\% of the participants choose a new strategy when confronted with a blocklist warning.
Only participants of the \fourDataSmallAbbr{} treatment with a very small blocklist, tended to keep their pre-warning strategy.
The edit distances vary slightly across the treatments and support this self-reported behavior:
participants in the 4-digit scenario changed on average 3 digits with the standard deviation showing that some participants changed their PIN completely while some participants only changed 2 digits.
The same conclusion can be drawn from the edit distances in the 6-digit case.

\begin{table}[t]
\vspace{-.2em}
\caption{Participants' PIN changing.}\label{tab:changing-strategies}
\vspace{-2em}
  \begin{center}
    \resizebox{1.0\columnwidth}{!}{\begin{tabular}{lr|rrrr|rr}
    \toprule
    \multicolumn{2}{c|}{}
    & \multicolumn{4}{c|}{\textbf{Selection vs.\ Changing Strategy}}
    & \multicolumn{2}{c}{\textbf{Edit Distance}} \\
    \textbf{Treatment}
    & \multicolumn{1}{c|}{\textbf{Hits}}
    & \multicolumn{1}{c}{\textbf{Sample}}
    & \multicolumn{1}{c}{\textbf{Same}}
    & \multicolumn{1}{c}{\textbf{Minor}}
    & \multicolumn{1}{c|}{\textbf{New}}
    & \multicolumn{1}{c}{\textbf{Mean}}
    & \multicolumn{1}{c}{\textbf{SD}} \\
    \midrule
    \fourPlaceboAbbr{}        & 122\:\: & 29 & 10 & 7 & 12 & 3.20 & 0.90 \\
    \fourIosWithAbbr{}        & $9^{\star}$   & 9  & 0  & 4 & 5  & 3.11 & 0.87 \\
    \fourIosNoAbbr{}        & 21\:\:  & 21 & 4  & 6 & 11 & 3.24 & 0.92 \\
    \fourDataSmallAbbr{}    & 5\:\:   & 5  & 2  & 2 & 1  & 3.20 & 0.75 \\
    \fourDataLargeAbbr{}    & 88\:\:  & 29 & 4  & 7 & 18 & 3.39 & 0.76 \\
    \midrule
    \sixPlaceboAbbr{}        & 117\:\: & 28 & 8  & 5 & 15 & 4.59 & 1.41 \\
    \sixIosWithAbbr{}        & $5^{\star}$   & 5  & 0  & 2 & 3  & 4.40 & 1.20 \\
    \bottomrule
    \end{tabular}}
\end{center}
    \begin{flushleft}
    {\footnotesize $\star$: Hit blocklist, and did not click-through.}
  \end{flushleft}
  \vspace{-3em}
\end{table}

%% file: 06-5-perception.tex
\subsection{User Perception}\label{sec:perception}
We analyzed participants' perceptions regarding PIN selections with respect to security and usability. 
Participants were asked to complete the phrase ``\emph{I feel the PIN I chose is}'' with three different adjectives: ``\emph{secure}, \emph{memorable}, and \emph{convenient}.''  The phrases were displayed randomly and participants responded using a Likert scale. The results are shown in Figure \ref{fig:perception-sec-mem-con}.
To compare these results, we converted the Likert responses into weighted averages on a scale of -2 to +2. As the weighted averages are not normally distributed, tested using the Shapiro-Wilk test ($p < 0.001$), we tested for initial differences using a Mann-Whitney $U$ test, followed with post-hoc, pair-wise tests using Dunn's-test comparisons of independent samples with a Bonferroni correction.

We found that there are significant differences across treatments when
considering Likert responses for {\em security}. Post-hoc analysis indicates
that the presence of a blocklist for 4-digit PINs increases the security
perception of the final PIN selected. This is supported by considering the
4-digit placebo treatment (\fourPlaceboAbbr{}) compared to the 4-digit control
(\fourControlAbbr{}). In the placebo treatment, every participant
interacted with a blocklist, and there is a significant increase in security
perceptions ($p<0.01$). We see similar differences for the large blocklist
treatment \fourDataLargeAbbr{} ($p<0.001$), where again, a large
portion (70\,\%) of participants encountered the blocklist. We did not see
significant differences for 6-digit PIN users after
encountering the blocklist. This may be because there is a pre-existing notion that
6-digit PINs are secure.

For {\em memorability} we also found significant differences
among the treatments. In post-hoc analysis we
found that increased interaction with the blocklist led to lower perceived
memorability of PINs, as evidenced by the \fourPlaceboAbbr{} ($p<0.001$),
\fourDataLargeAbbr{} ($p < 0.001$), and the \sixPlaceboAbbr{} ($p<0.01$) treatments compared to their
respective control treatments. The \fourDataLargeAbbr{} showed the most
significant differences with other treatments, likely due to the fact that many
participants encountered the blocklist for multiple PIN choices and thus were
relying on not just second-choice PINs, but also third- and fourth-choice, etc. PINs
that are perceived to be less memorable.

The responses to perceived {\em convenience} also show significant differences, however, post-hoc analysis revealed limited effects when considering
pair-wise comparisons. In general, participants perceived their 4-digit PINs at
the same convenience level across treatments. While we did not see a significant
difference between convenience levels between 4- and 6-digit PINs, the perceived
convenience of 6-digit PINs may be precarious because we observed a
significant difference ($p<0.01$) between the 6-digit placebo and control
treatments. This suggests that while a user may be comfortable with their
first-choice 6-digit PIN, there is much higher perceived {\em inconvenience}
for their second-choice 6-digit PIN.

%% file: 06-6-sentiment.tex
\subsection{User Sentiment}\label{sec:sentiment}
To gain insight into participants' sentiments regarding blocking, we asked
``\emph{Please describe three general feelings or reactions that you had after you received this warning message}'' or ``{\em would have had''} if the participant did not encounter a blocklist.  Accompanying the prompt are three free-form, short text
fields.  A codebook was constructed by two individual coders summarized in
Appendix~\ref{app:feelings} in Table \ref{tab:feelings}. 
For each of the four
categories (blocklist hit experienced vs. imagined, 4- vs. 6-digit PINs, non-enforcing vs. enforcing, different blocklist sizes), 21 individuals' responses were randomly selected.
Again, two individual raters were tasked with coding the responses. The
inter-rater reliability, computed using Cohen's kappa, was $\kappa = 0.91$.

Using the NRC Word-Emotion Association Lexicon~\cite{mohammad-13-nrc-database},
we classified assigned codes in terms of sentiment (positive, negative, or
neutral) for Figure \ref{fig:sentiment}.  EmoLex maps
individual English words (in this case, codes assigned by our coders) to exactly
one sentiment.  For example, ``indifference,'' is labeled with the ``negative''
sentiment.  As expected, participants generally had a negative reaction to the
blocklist warning message.

\input{fig/plots/perception/sentiment}

While overall, participants expressed negative sentiments towards blocklist
messages, which may be expected as warning messages are not often well received
by users~\cite{akhawe-13-alice}, we only observed significant differences in a
single comparison. 

Using a $\chi^2$ test, we found that there was significant difference ($p<0.05$)
in the proportion of negative sentiment when considering PIN length for the two
placebo treatments. As both groups always experienced a blocklist event, a higher
negative sentiment exists for the placebo blocklist with 4-digits. This might be
because users were confused and angered by the warning as the blocklist event
was arbitrary. However, in the 6-digit PIN case, less familiarity with 6-digit PINs may have led to less negative reactions.

Interestingly, participants in general consider displaying warnings about weak PIN choices to be appropriate although they cannot imagine that their own choice might be considered insecure.
Moreover, sentiments are similar for those who~hit the blocklist and those who imagined having done so. 
This suggests that future research on blocklist warning design may benefit from simply asking participants to imagine such events.

%% file: fig/plots/perception/sentiment.tex
\definecolor{pos}{HTML}{ABD4B4} 
\definecolor{neu}{HTML}{A0A0A0} 
\definecolor{neg}{HTML}{EF959A} 

\begin{figure}[t]
  \centering
\resizebox{0.74\columnwidth}{!}{
\small
\renewcommand*{\arraystretch}{0.7}
    \centering
    \setlength{\tabcolsep}{0pt}
    \begin{tabular}{r c}
        \toprule
        \textbf{Group} & \textbf{Sentiment}\\
        \midrule

& Experienced vs. Imagined Blocklist \\
\midrule

\raisebox{0.3em}{Experienced} & \scalebox{0.7}{
\begin{tikzpicture}
\begin{axis}[xbar stacked,bar width=1cm,ytick=\empty,y post scale=0.07,xmin=0,xmax=100,ymin=0,ymax=1,legend style={at={(0,-3)},
legend style={cells={align=left, anchor=center, fill}, nodes={inner sep=0.4ex,below=-2ex}},
column sep=0cm, draw=none, anchor=west, legend columns=3},xticklabel style={opacity=0,yshift=12pt}]
\addplot[fill=neg, xbar legend, mark=none] coordinates {(68.33,1)};
\addplot[fill=neu, xbar legend, mark=none] coordinates {(16.67,1)};
\addplot[fill=pos, xbar legend, mark=none] coordinates {(15.0,1)};
\end{axis}
\end{tikzpicture}}\\

\raisebox{0.3em}{Imagined} & \scalebox{0.7}{
\begin{tikzpicture}
\begin{axis}[xbar stacked,bar width=1cm,ytick=\empty,y post scale=0.07,xmin=0,xmax=100,ymin=0,ymax=1,legend style={at={(0.15,-2)},
legend style={cells={align=left, anchor=center, fill}, nodes={inner sep=0.4ex,below=-2ex}},
column sep=0cm, draw=none, anchor=west, legend columns=3},xticklabel style={opacity=0,yshift=12pt}]
\addplot[fill=neg, xbar legend, mark=none] coordinates {(64.52,1)};
\addplot[fill=neu, xbar legend, mark=none] coordinates {(16.13,1)};
\addplot[fill=pos, xbar legend, mark=none] coordinates {(19.35,1)};
\end{axis}
\end{tikzpicture}}\\

\midrule
& PIN Length \\
\midrule

\raisebox{0.3em}{\fourPlaceboAbbr} & \scalebox{0.7}{
\begin{tikzpicture}
\begin{axis}[xbar stacked,bar width=1cm,ytick=\empty,y post scale=0.07,xmin=0,xmax=100,ymin=0,ymax=1,legend style={at={(0.15,-2)},
legend style={cells={align=left, anchor=center, fill}, nodes={inner sep=0.4ex,below=-2ex}},
column sep=0cm, draw=none, anchor=west, legend columns=3},xticklabel style={opacity=0,yshift=12pt}]
\addplot[fill=neg, xbar legend, mark=none] coordinates {(74.19,1)};
\addplot[fill=neu, xbar legend, mark=none] coordinates {(11.29,1)};
\addplot[fill=pos, xbar legend, mark=none] coordinates {(14.52,1)};
\end{axis}
\end{tikzpicture}}\\

\raisebox{0.3em}{\sixPlaceboAbbr} & \scalebox{0.7}{
\begin{tikzpicture}
\begin{axis}[xbar stacked,bar width=1cm,ytick=\empty,y post scale=0.07,xmin=0,xmax=100,ymin=0,ymax=1,legend style={at={(0.15,-2)},
legend style={cells={align=left, anchor=center, fill}, nodes={inner sep=0.4ex,below=-2ex}},
column sep=0cm, draw=none, anchor=west, legend columns=3},xticklabel style={opacity=0,yshift=12pt}]
\addplot[fill=neg, xbar legend, mark=none] coordinates {(52.46,1)};
\addplot[fill=neu, xbar legend, mark=none] coordinates {(24.59,1)};
\addplot[fill=pos, xbar legend, mark=none] coordinates {(22.95,1)};
\end{axis}
\end{tikzpicture}}\\

\midrule
& Non-Enforcing vs.\ Enforcing\\
\midrule

\raisebox{0.3em}{\fourIosWithAbbr} & \scalebox{0.7}{
\begin{tikzpicture}
\begin{axis}[xbar stacked,bar width=1cm,ytick=\empty,y post scale=0.07,xmin=0,xmax=100,ymin=0,ymax=1,legend style={at={(0.15,-2)},
legend style={cells={align=left, anchor=center, fill}, nodes={inner sep=0.4ex,below=-2ex}},
column sep=0cm, draw=none, anchor=west, legend columns=3},xticklabel style={opacity=0,yshift=12pt}]
\addplot[fill=neg, xbar legend, mark=none] coordinates {(65,1)};
\addplot[fill=neu, xbar legend, mark=none] coordinates {(15,1)};
\addplot[fill=pos, xbar legend, mark=none] coordinates {(20,1)};
\end{axis}
\end{tikzpicture}}\\

\raisebox{0.3em}{\fourIosNoAbbr} & \scalebox{0.7}{
\begin{tikzpicture}
\begin{axis}[xbar stacked,bar width=1cm,ytick=\empty,y post scale=0.07,xmin=0,xmax=100,ymin=0,ymax=1,legend style={at={(0.15,-2)},
legend style={cells={align=left, anchor=center, fill}, nodes={inner sep=0.4ex,below=-2ex}},
column sep=0cm, draw=none, anchor=west, legend columns=3},xticklabel style={opacity=0,yshift=12pt}]
\addplot[fill=neg, xbar legend, mark=none] coordinates {(50.82,1)};
\addplot[fill=neu, xbar legend, mark=none] coordinates {(18.03,1)};
\addplot[fill=pos, xbar legend, mark=none] coordinates {(31.15,1)};
\end{axis}
\end{tikzpicture}}\\

\midrule
& Blocklist Size\\
\midrule

\raisebox{0.3em}{\fourPlaceboAbbr{}} & \scalebox{0.7}{
\begin{tikzpicture}
\begin{axis}[xbar stacked,bar width=1cm,ytick=\empty,y post scale=0.07,xmin=0,xmax=100,ymin=0,ymax=1,legend style={at={(0.15,-2)},
legend style={cells={align=left, anchor=center, fill}, nodes={inner sep=0.4ex,below=-2ex}},
column sep=0cm, draw=none, anchor=west, legend columns=3},xticklabel style={opacity=0,yshift=12pt}]
\addplot[fill=neg, xbar legend, mark=none] coordinates {(74.19,1)};
\addplot[fill=neu, xbar legend, mark=none] coordinates {(11.29,1)};
\addplot[fill=pos, xbar legend, mark=none] coordinates {(14.52,1)};
\end{axis}
\end{tikzpicture}}\\

\raisebox{0.3em}{\fourIosNoAbbr{}} & \scalebox{0.7}{
\begin{tikzpicture}
\begin{axis}[xbar stacked,bar width=1cm,ytick=\empty,y post scale=0.07,xmin=0,xmax=100,ymin=0,ymax=1,legend style={at={(0.15,-2)},
legend style={cells={align=left, anchor=center, fill}, nodes={inner sep=0.4ex,below=-2ex}},
column sep=0cm, draw=none, anchor=west, legend columns=3},xticklabel style={opacity=0,yshift=12pt}]
\addplot[fill=neg, xbar legend, mark=none] coordinates {(50.82,1)};
\addplot[fill=neu, xbar legend, mark=none] coordinates {(18.03,1)};
\addplot[fill=pos, xbar legend, mark=none] coordinates {(31.15,1)};
\end{axis}
\end{tikzpicture}}\\

\raisebox{2.5em}{\fourDataLargeAbbr{}} & \scalebox{0.7}{
\begin{tikzpicture}
\begin{axis}[xbar stacked,bar width=1cm,ytick=\empty,y post scale=0.07,xmin=0,xmax=100,ymin=0,ymax=1,legend style={at={(0.15,-2)},
legend style={cells={align=left, anchor=center, fill}, nodes={inner sep=0.4ex,below=-2ex}},
column sep=0cm, draw=none, anchor=west, legend columns=3}]
\addplot[fill=neg, xbar legend, mark=none] coordinates {(64.52,1)};
\addplot[fill=neu, xbar legend, mark=none] coordinates {(11.29,1)};
\addplot[fill=pos, xbar legend, mark=none] coordinates {(24.19,1)};
\addlegendentry{{\normalsize Negative}\hspace{0.5em}};
\addlegendentry{{\normalsize Neutral}\hspace{0.5em}};
\addlegendentry{{\normalsize Positive}\hspace{0.5em}};
\end{axis}
\end{tikzpicture}}\\

    \bottomrule
    \end{tabular}}
    \caption{Participants' sentiment: We split the participants into four categories and classified their feelings in terms of sentiment using EmoLex~\cite{mohammad-13-nrc-database}.}\label{fig:sentiment}
    \vspace{-.2in}
\end{figure}
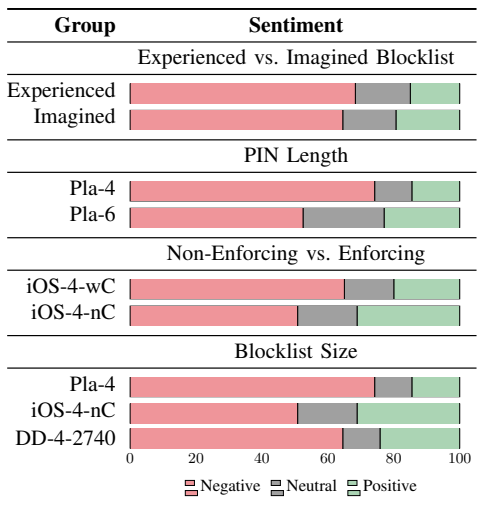 

%% file: 07-conclusion.tex
\section{Conclusion and Recommendations}
\label{sec:conclusion}
This paper presents the first comprehensive study of PIN security as primed for the smartphone unlock setting.  In the smartphone unlock setting, developers have adopted notable countermeasures---throttling, blocking, PIN length---which we consider as part of our analysis. Using a throttled attacker model, we find that \mbox{6-digit} PINs offer little to no advantage, and sometimes make matters worse. Also, blocklists would have to be far larger than those in use on today's mobile operating systems to affect security.

Given this information, we offer a number of recommendations to
mobile developers.
\begin{itemize}[noitemsep,nolistsep]
\item In a throttled scenario, simply increasing the PIN length is of little benefit.
In our results, there was no significant difference between 4- and 6-digit PINs within the first 100 guesses.
To justify the adoption of longer PINs, developers should carefully articulate an alternative threat model. Observe that without throttling, an attacker could quickly try all 4- and 6-digit PINs.

\item On iOS, with only 10 possible guesses, we could not observe any security benefits when a blocklist is deployed, either for 4- or 6-digit PINs.
On Android, where 100 guesses are feasible, we find that a blocklist would be beneficial for 4-digit PINs. However, such a blocklist would need to contain roughly 10\% of the PIN space which is much more than currently deployed blocklists. More research is needed to test the effectiveness of blocklists for 6-digit PINs.

\item We observe that the perceived convenience is lower when users are forced to select a second 6-digit PIN as compared to selecting a second 4-digit PIN (as was the case in the placebo treatments). This may suggest users are less familiar with selecting 6-digit PINs, but the reasons for this are left to future investigation.

\item While we observed advantages for using a placebo blocklist in the unthrottled settings, we do not recommend implementing a placebo blocklist, as users will simply game it once the deception is known.
\end{itemize}

%% file: 97-appendix.tex
\section{Appendix}\label{sec:appendix}

\subsection{Survey Instrument}\label{app:survey}
\input{app/97a-survey-instrument}
\clearpage

\onecolumn

\subsection{Demographics}\label{app:demographics}
\input{app/97b-demographics}

\subsection{Device Usage}\label{app:device-usage}
\input{app/97c-device-usage}
\clearpage

\subsection{Feelings and Sentiments}\label{app:feelings}
\input{app/97d-qualitative}

\subsection{PIN Selection and Changing Strategies}\label{app:selectiong-changing}
\input{app/97e-selection-changing.tex}
\clearpage

%% file: app/97a-survey-instrument.tex
\definecolor{structure}{HTML}{03588C} 
\definecolor{note}{HTML}{BF2C47} 
\newcommand{\questionspace}{\vspace{1em}}

\mbox{}
\setlength\parindent{0pt}
\begin{footnotesize}
\\
\textcolor{structure}{Questions for participants who \textbf{hit} the blocklist.}

\vspace{.5em}

We noticed that you received the following warning while choosing your PIN:

\questionspace

\textit{{\color{note} [A screenshot of the same warning message that the participant saw during the study.]}}

\questionspace

People use different strategies for choosing their PINs. Below, we will ask about your strategy.

\questionspace

\begin{enumerate}
\item Prior to seeing the warning above, what was your strategy for choosing your PIN? \newline
Answer: \rule{3.5cm}{.1pt}

\questionspace

\item After receiving the warning message, please describe how or if your strategy changed when choosing your PIN. \newline
Answer: \rule{3.5cm}{.1pt}

\vspace{1em}

\textit{{\color{note} The ``Extra''  question was only asked if the participant had the option to ignore the warning and did so by clicking ``Use Anyway.''}}

\vspace{.25em}

\item[(Extra)] You selected ``Use Anyway'' when choosing your final PIN. Please describe why you did not change your final PIN after seeing this warning message.  \newline
Answer: \rule{3.5cm}{.1pt}

\questionspace

\item Please describe three general feelings or reactions that you had after you received this warning message. \newline
Feeling 1: \rule{1.4cm}{.1pt} Feeling 2: \rule{1.4cm}{.1pt} Feeling 3: \rule{1.4cm}{.1pt}
\end{enumerate}

\questionspace

Please select the answer choice that most closely matches how you feel about the following statements:

\questionspace

\begin{enumerate}
\setcounter{enumi}{3}
\item My initial PIN creation strategy caused the display of this warning. \newline
$\circ$~Strongly agree
$\circ$~Agree
$\circ$~Neutral
$\circ$~Disagree
$\circ$~Strongly Disagree
\end{enumerate}

\questionspace

\textcolor{structure}{Questions for participants who did \textbf{not hit} the blocklist.}
\vspace{0.5em}\\
People use different strategies for choosing their PINs. Below, we will ask about your strategy.

\questionspace

\begin{enumerate}
\setcounter{enumi}{0}
\item What was your strategy for choosing your PIN? \newline
Answer: \rule{3.5cm}{.1pt}

\questionspace

Imagine you received the following warning message after choosing your PIN:

\textit{{\color{note} \newline[A screenshot of the warning message as in Figure~\ref{fig:feedback:ct} or Figure~\ref{fig:feedback:noct}.]\\}}

\item Please describe how or if your strategy would change as a result of the message. \newline
Answer: \rule{3.5cm}{.1pt}

\questionspace

\item Please describe three general feelings or reactions that you would have had after you received this warning message. \newline
Feeling 1: \rule{1.4cm}{.1pt} Feeling 2: \rule{1.4cm}{.1pt} Feeling 3: \rule{1.4cm}{.1pt}
\end{enumerate}

\questionspace

Please select the answer choice that most closely matches how you feel about the following statements:

\questionspace

\begin{enumerate}
\setcounter{enumi}{3}
\item My PIN creation strategy would cause this warning message to appear. \newline
$\circ$~Strongly agree
$\circ$~Agree
$\circ$~Neutral
$\circ$~Disagree
$\circ$~Strongly Disagree
\end{enumerate}

\questionspace

\textcolor{structure}{From now on all participants saw the same questions.}

\vspace{.5em}

\begin{enumerate}
\setcounter{enumi}{4}
\item It is appropriate for smartphones to display warning messages about PIN security. \newline
$\circ$~Strongly agree
$\circ$~Agree
$\circ$~Neutral
$\circ$~Disagree
$\circ$~Strongly Disagree
\end{enumerate}
\newpage
Please select the answer choice that most closely matches how you feel about the following statements referring to the final PIN you chose:

\vspace{1em}

\begin{quote}
\textit{{\color{note} The order of questions 6, 7, and 9 was chosen randomly for each participant. The attention check question was always the 8th question.}}
\end{quote}

\vspace{0em}

\begin{enumerate}
\setcounter{enumi}{5}
\item I feel the PIN I chose is: \newline
$\circ$~Secure
$\circ$~Somewhat secure
$\circ$~Neither easy nor insecure
$\circ$~Somewhat insecure
$\circ$~Insecure

\vspace{.875em}

\item I feel the PIN I chose is: \newline
$\circ$~Easy to remember
$\circ$~Somewhat easy to remember
$\circ$~Neither easy nor hard to remember
$\circ$~Somewhat hard to remember
$\circ$~Difficult to remember

\vspace{.875em}

\item What is the shape of a red ball? \newline
$\circ$~Red
$\circ$~Blue
$\circ$~Square
$\circ$~Round

\vspace{.85em}

\item I feel the PIN I chose is: \newline
$\circ$~Easy to enter
$\circ$~Somewhat easy to enter
$\circ$~Neither easy nor hard to enter
$\circ$~Somewhat hard to enter
$\circ$~Difficult to enter

\vspace{.85em}

\item What is your age range? \newline
$\circ$~18-24
$\circ$~25-34
$\circ$~35-44
$\circ$~45-54
$\circ$~55-64
$\circ$~65-74
$\circ$~75 or older \\
$\circ$~Prefer not to say

\vspace{.85em}

\item With what gender do you identify? \newline
$\circ$~Male
$\circ$~Female
$\circ$~Non-Binary
$\circ$~Other
$\circ$~Prefer not to say

\vspace{.85em}

\item What is the highest degree or level of school you have completed? \newline
$\circ$~Some high school
$\circ$~High school
$\circ$~Some college
$\circ$~Trade, technical, or vocational training
$\circ$~Associate's Degree
$\circ$~Bachelor's Degree
$\circ$~Master's Degree
$\circ$~Professional Degree
$\circ$~Doctorate
$\circ$~Prefer not to say

\vspace{.875em}

\item Do you use any of the following biometrics to unlock your primary smartphone? (Select all that apply) \label{item:biometric} \newline
$\square$ Fingerprint
$\square$~Face
$\square$~Iris
$\square$~Other biometric
$\square$~I do not use a biometric
$\square$~I do not use a smartphone
$\square$~Prefer not to say

\vspace{.875em}

\textit{{\color{note} If the participant stated they use a biometric in question \ref{item:biometric}:}}

\vspace{0.25em}

\item[14A)] How do you unlock your smartphone, if your biometric fails or when you reboot your primary smartphone? \newline
$\circ$~None
$\circ$~Pattern
$\circ$~4-digit PIN
$\circ$~6-digit PIN
$\circ$~PIN of other length
$\circ$~Alphanumeric password
$\circ$~I use an unlock method not listed here \\
$\circ$~I do not use a smartphone
$\circ$~Prefer not to say

\vspace{.875em}

\textit{{\color{note} If the participant stated they do not use a biometric in question \ref{item:biometric}:}}

\vspace{0.25em}

\item[(14B)] What screen lock do you use to unlock your primary smartphone? \newline
$\circ$~None
$\circ$~Pattern
$\circ$~4-digit PIN
$\circ$~6-digit PIN
$\circ$~PIN of other length
$\circ$~Alphanumeric password
$\circ$~I use an unlock method not listed here \\
$\circ$~I do not use a smartphone
$\circ$~Prefer not to say

\vspace{.875em}

\setcounter{enumi}{14}
\item What is the operating system of your primary smartphone? \newline
$\circ$~Android
$\circ$~iOS (iPhone)
$\circ$~Other
$\circ$~I do not use a smartphone \\
$\circ$~Prefer not to say

\vspace{.875em}

\item Which of the following best describes your educational background or job field? \newline
$\circ$~I have an education in, or work in, the field of computer science, computer engineering or IT. \newline
$\circ$~I do not have an education in, nor do I work in, the field of computer science, computer engineering or IT. \newline
$\circ$~Prefer not to say to say

\vspace{.875em}

\item Please indicate if you have honestly participated in this survey and followed instructions completely. You will not be penalized/rejected for indicating 'No' but your data may not be included in the analysis: \newline
$\circ$~Yes
$\circ$~No

\vspace{.875em}

\item Please feel free to provide any final feedback you may have in the field below. \newline
Answer: \rule{3.5cm}{.1pt}

\end{enumerate}

\end{footnotesize}

%% file: app/97b-demographics.tex
\vspace{-1em}
\begin{table*}[h]
  \centering
\caption{Overall demographics of the participants. For the sake of clarity, we grouped answers for \textit{Non-Binary}, \textit{Other}, and \textit{Prefer not to say} under \textit{Other}.}\label{tab:demographics}
\resizebox{0.73\textwidth}{!}{\begin{tabular}{rrrrrrrrr} \toprule
 & \multicolumn{2}{c}{\textbf{Male}} 
 & \multicolumn{2}{c}{\textbf{Female}} 
 & \multicolumn{2}{c}{\textbf{Other}} 
 & \multicolumn{2}{c}{\textbf{Total}} \\    
 \cmidrule(lr){2-3} \cmidrule(lr){4-5} \cmidrule(lr){6-7} \cmidrule(lr){8-9}
 & \multicolumn{1}{r}{\textbf{No.}} & \multicolumn{1}{c}{\textbf{\%}} 
 & \multicolumn{1}{r}{\textbf{No.}} & \multicolumn{1}{c}{\textbf{\%}} 
 & \multicolumn{1}{r}{\textbf{No.}} & \multicolumn{1}{c}{\textbf{\%}} 
 & \multicolumn{1}{r}{\textbf{No.}} & \multicolumn{1}{c}{\textbf{\%}} \\
\midrule
\multirow{1}{10cm}{\textbf{What is your age range?}} & 619 & 51\,\% & 590 & 48\,\% & 11 & 1\,\% & 1220 & 100\,\% \\
\midrule
18--24 & 85     & 7\,\%     & 76    & 6\,\%     & 4 & 0\,\% &   165 & 13\,\% \\
25--34 & 309    & 25\,\%    & 267   & 22\,\%    & 4 & 0\,\% &   580 & 47\,\% \\
35--44 & 147    & 12\,\%    & 145   & 12\,\%    & 2 & 0\,\% &   294 & 24\,\% \\
45--54 & 56     & 5\,\%     & 63    & 5\,\%     & 0 & 0\,\% &   119 & 10\,\% \\
55--64 & 16     & 1\,\%     & 35    & 3\,\%     & 0 & 0\,\% &   51  & 4\,\%  \\
65--74 & 6      & 1\,\%     & 4     & 0\,\%     & 0 & 0\,\% &   10  & 1\,\%  \\
Prefer not to say   & 0 & 0\,\% & 0 & 0\,\% & 1 & 0\,\% & 1 & 0\,\%  \\
\midrule
\multirow{1}{10cm}{\textbf{What is the highest degree or level of school you have completed?}} & 619 & 51\,\% & 590 & 48\,\% & 11 & 1\,\% & 1220 & 100\,\% \\
\midrule
Some High School    & 2   & 0\,\%   & 3     & 0\,\%     & 0 & 0\,\% &   5 & 0\,\% \\
High School         & 63  & 5\,\%   & 52    & 4\,\%     & 2 & 0\,\% &   117 & 10\,\% \\
Some College        & 154 & 13\,\%  & 116   & 10\,\%    & 5 & 0\,\% &   275 & 23\,\% \\
Training            & 23  & 2\,\%   & 23    & 2\,\%     & 0 & 0\,\% &   46 & 4\,\% \\
Associates          & 63  & 5\,\%   & 82    & 7\,\%     & 1 & 0\,\% &   146 & 12\,\%  \\
Bachelor's          & 236 & 19\,\%  & 235   & 19\,\%    & 2 & 0\,\% &   473 & 39\,\%  \\
Master's            & 54  & 5\,\%   & 66    & 5\,\%     & 0 & 0\,\% &   120 & 9\,\%  \\
Professional        & 11  & 1\,\%   & 4     & 0\,\%     & 0 & 0\,\% &   15 & 1\,\%  \\
Doctorate           & 12  & 1\,\%   & 9     & 1\,\%     & 0 & 0\,\% &   21 & 2\,\%  \\
Prefer not to say   & 1   & 0\,\%   & 0     & 0\,\%     & 1 & 0\,\% &   2 & 0\,\%  \\
\midrule
\multirow{1}{10cm}{\textbf{Which of the following best describes your educational background or job field?}} & 619 & 51\,\% & 590 & 48\,\% & 11 & 1\,\% & 1220 & 100\,\% \\
\midrule
Tech                & 231 & 30\,\%  & 83  & 7\,\%   & 3 & 0\,\% &   317 & 26\,\% \\
No Tech             & 368 & 19\,\%  & 491 & 40\,\%  & 7 & 1\,\% &   866 & 71\,\% \\
Prefer not to say   & 20  & 2\,\%   & 16  & 1\,\%   & 1 & 0\,\% &   37  & 3\,\% \\
\bottomrule
\end{tabular}}
\end{table*}

%% file: app/97c-device-usage.tex
\vspace{-1em}
\begin{table*}[h]
\centering
\caption{Answers of participants regarding their device usage. Note, for the biometrics question, participants selected all that apply. For the sake of clarity, we grouped answers for \textit{Non-Binary}, \textit{Other}, and \textit{Prefer not to say} under \textit{Other}.}\label{tab:device-usage}
\resizebox{0.9\textwidth}{!}{\begin{tabular}{rrrrrrrrr} \toprule
 & \multicolumn{2}{c}{\textbf{Male}}
 & \multicolumn{2}{c}{\textbf{Female}}
 & \multicolumn{2}{c}{\textbf{Other}}
 & \multicolumn{2}{c}{\textbf{Total}} \\
 \cmidrule(lr){2-3} \cmidrule(lr){4-5} \cmidrule(lr){6-7} \cmidrule(lr){8-9}
 & \multicolumn{1}{r}{\textbf{No.}} & \multicolumn{1}{c}{\textbf{\%}}
 & \multicolumn{1}{r}{\textbf{No.}} & \multicolumn{1}{c}{\textbf{\%}}
 & \multicolumn{1}{r}{\textbf{No.}} & \multicolumn{1}{c}{\textbf{\%}}
 & \multicolumn{1}{r}{\textbf{No.}} & \multicolumn{1}{c}{\textbf{\%}} \\
\midrule
\multirow{1}{14cm}{\textbf{Do you use any of the following biometrics to unlock your primary smartphone?}} & 619 & 51\,\% & 590 & 48\,\% & 11 & 1\,\% & 1220 & 100\,\% \\
\midrule
Fingerprint     & 329   & 27\,\%    & 310   & 25\,\%    & 7 & 1\,\% & 646 & 53\,\% \\
Face            & 95    & 8\,\%     & 67    & 5\,\%     & 0 & 0\,\% & 162 & 13\,\% \\
Iris            & 24    & 2\,\%     & 11    & 1\,\%     & 0 & 0\,\% & 35    & 3\,\% \\
Other Biometric & 13    & 1\,\%     & 15    & 1\,\%     & 0 & 0\,\% & 28 & 2\,\% \\
No Biometric    & 209   & 17\,\%    & 206   & 17\,\%    & 3 & 0\,\% & 418  & 34\,\%  \\
No Smartphone   & 1     & 0\,\%     & 0     & 0\,\%     & 0 & 0\,\% & 1  & 0\,\%  \\
Prefer not to say   & 19 & 2\,\%    & 22    & 2\,\%     & 1 & 0\,\% & 42 & 3\,\%  \\
\midrule
\multirow{1}{14cm}{\textbf{How do you unlock your smartphone, if your biometric fails or when you reboot your primary smartphone?}} & 390 & 51\,\% & 362 & 48\,\% & 7 & 1\,\% & 759 & 100\,\% \\
\midrule
None                & 2   & 0\,\%   & 5     & 1\,\%     & 0 & 0\,\% &   7 & 1\,\% \\
Pattern             & 65  & 8\,\%   & 48    & 6\,\%     & 0 & 0\,\% &   113 & 15\,\% \\
4-digit PIN         & 183 & 24\,\%  & 186   & 25\,\%    & 3 & 0\,\% &   372 & 49\,\% \\
6-digit PIN         & 98  & 13\,\%  & 99    & 13\,\%    & 4 & 1\,\% &   201 & 26\,\% \\
PIN of other length & 12  & 2\,\%   & 10    & 1\,\%     & 0 & 0\,\% &   22 & 3\,\%  \\
Alphanumeric        & 21  & 3\,\%   & 11    & 2\,\%     & 0 & 0\,\% &   32 & 4\,\%  \\
Other method        & 6   & 1\,\%   & 2     & 0\,\%     & 0 & 0\,\% &   8 & 1\,\%  \\
No smartphone       & 0   & 0\,\%   & 0     & 0\,\%     & 0 & 0\,\% &   0 & 0\,\%  \\
Prefer not to say   & 3   & 0\,\%   & 1     & 0\,\%     & 0 & 0\,\% &   4 & 1\,\%  \\
\midrule
\multirow{1}{14cm}{\textbf{What screen lock do you use to unlock your primary smartphone?}} & 229 & 50\,\% & 228 & 50\,\%   & 4 & 0\,\% & 461 & 100\,\% \\
\midrule
None                & 58 & 13\,\%   & 82 & 18\,\%   & 0 & 0\,\% &   140 & 30\,\% \\
Pattern             & 36 & 8\,\%    & 23 & 5\,\%    & 1 & 0\,\% &   60 & 13\,\% \\
4-digit PIN         & 83 & 18\,\%   & 81 & 18\,\%   & 1 & 0\,\% &   165 & 36\,\% \\
6-digit PIN         & 20 & 4\,\%    & 17 & 4\,\%    & 0 & 0\,\% &   37 & 8\,\% \\
PIN of other length & 6  & 1\,\%    & 2  & 1\,\%    & 0 & 0\,\% &   8 & 2\,\%  \\
Alphanumeric        & 6  & 1\,\%    & 6  & 1\,\%    & 0 & 0\,\% &   12 & 3\,\%  \\
Other method        & 7  & 2\,\%    & 4  & 1\,\%    & 0 & 0\,\% &   11 & 2\,\%  \\
No smartphone       & 0  & 0\,\%    & 1  & 0\,\%    & 0 & 0\,\% &   1 & 0\,\%  \\
Prefer not to say   & 13 & 3\,\%    & 12 & 2\,\%    & 2 & 1\,\% &   27 & 6\,\%  \\
\midrule
\multirow{1}{14cm}{\textbf{What is the operating system of your primary smartphone?}} & 619 & 51\,\% & 590 & 48\,\% & 11 & 1\,\% & 1220 & 100\,\% \\
\midrule
Android             & 382 & 31\,\%  & 310  & 25\,\%     & 6 & 1\,\% &   698 & 57\,\% \\
iOS                 & 232 & 19\,\%  & 270 & 22\,\%  & 4 & 1\,\% &   506 & 42\,\% \\
Other               & 1 & 0\,\%     & 4 & 0\,\%     & 0 & 0\,\% &   5 & 0\,\% \\
No smartphone       & 0 & 0\,\%     & 0 & 0\,\%     & 0 & 0\,\% &   0 & 0\,\% \\
Prefer not to say   & 4  & 1\,\%    & 6  & 1\,\%    & 1 & 0\,\% &   11  & 1\,\% \\
\bottomrule
\end{tabular}}
\end{table*}

%% file: app/97d-qualitative.tex
\definecolor{pos}{HTML}{357F2E} 
\definecolor{neu}{HTML}{737373} 
\definecolor{neg}{HTML}{CE2029} 
\vspace{-1em}
\begin{table*}[h]
\caption{As part of our questionnaire, we asked participants for 3~feelings about the blocklist warning. We coded and analyzed these feelings from a sample of 130~participants that encountered a blocklist. We also included 21~participants that only imagined hitting a blocklist. Below, we list the top~20 reported feelings. Two coders independently coded the data and the level of agreement between the coders, measured by Cohen's kappa was $\kappa = 0.91$. Question: ``\textit{Please describe three general feelings or reactions that you had after you received this warning message.''} or ``\textit{Please describe three general feelings or reactions that you would have had after you received this warning message.''}}
\label{tab:feelings}
\centering
\resizebox{0.5\textwidth}{!}{\begin{tabular}{@{}rcll@{}}
\toprule
\textbf{Code Name} & \textbf{Frequency} & \textbf{Sample from the Study} & \textbf{Sentiment}\\
\midrule
Annoyance       & 92    & ``Annoyed by this message.'' & {\color{neg} Negative} \\
Frustrated      & 45    & ``This message frustrates me.'' & {\color{neg} Negative} \\
Worried         & 41    & ``I am worried about my PIN's security.'' & {\color{neg} Negative} \\
Indifference    & 34    & ``Don't care about this message.'' & {\color{neg} Negative} \\
Surprised       & 32    & ``Surprised to see this message.'' & {\color{neu} Neutral} \\
Fear            & 32    & ``Afraid of attackers.'' & {\color{neg} Negative} \\
Doubt           & 32    & ``I distrust the veracity of this message.'' & {\color{neg} Negative} \\
Thinking        & 31    & ``Thinking about my PIN's security.'' & {\color{neu} Neutral} \\
Acceptance      & 27    & ``I agree with this message.'' & {\color{pos} Positive} \\
Compelling      & 26    & ``Motivated to change my PIN.'' & {\color{pos} Positive} \\
Cautious        & 26    & ``Cautious about my PIN.'' & {\color{pos} Positive} \\
Confusion       & 22    & ``This message is confusing.'' & {\color{neg} Negative} \\
Happy           & 19    & ``Happy my PIN will be stronger.'' & {\color{pos} Positive} \\
Shame           & 18    & ``Ashamed my PIN wasn't strong.'' & {\color{neg} Negative} \\
Remember        & 13    & ``I might forget my PIN.'' & {\color{neu} Neutral} \\
Angry           & 13    & ``Angry this message appeared.'' & {\color{neg} Negative} \\
Curiosity       & 12    & ``I wonder why this message appeared.'' & {\color{pos} Positive} \\
Alert           & 10    & ``I'm now more aware.'' & {\color{neu} Neutral} \\
Safe            & \,~8  & ``Confident this PIN will be safe.'' & {\color{pos} Positive} \\
Sadness         & \,~7  & ``Sad this message appeared.'' & {\color{neg} Negative} \\
\bottomrule
\end{tabular}}
\end{table*}

%% file: app/97e-selection-changing.tex
\vspace{-1em}
\begin{table}[htbp]
  \caption{We coded and analyzed a sample of 200~PIN selection strategies. Below, we list the top~10 selection strategies. Two coders independently coded the data. The level of agreement among the coders, measured by Cohen's kappa, was $\kappa = 0.92$. Question: ``\emph{People use different strategies for choosing their PINs. Below, we will ask about your strategy. What was your strategy for choosing your PIN?}''}\label{tab:selection-strategies}%
  \centering
    \resizebox{0.9\columnwidth}{!}{\begin{tabular}{rclcl}
    \toprule
    \textbf{Code Name} & \textbf{Frequency} & \textbf{Description} & \textbf{Example PIN} & \textbf{Sample from the Study} \\
    \midrule
    Date  & 59    & Special date like anniversary, birthday, graduation day & 1987 / 112518 & ``A date I won't forget.'' \\
    Memorable & 37    & Memorability was the main concern & 2827 / 777888 & ``A number easy to remember.'' \\
    Pattern & 24    & Visualized a pattern on the PIN pad & 2580 / 137955 & ``The numbers on how they appeared on the PIN pad.'' \\
    Meaning & 20    & Personal meaning; Familiar or significant number & 6767 / 769339 & ``I chose my favorite numbers and used them repeatedly.'' \\
    Random & 14    & Randomly chosen digits & 4619 / 568421 & ``Random numbers that do not repeat.'' \\
    Reuse & 12    & Reused PIN from a different device/service & 0596 / 260771 & ``The one I normally use.'' \\
    Word  & \,~9     & Textonyms; Converted a word to a number & 2539 / 567326 & ``Dog name.'' \\
    Simple & \,~9     & Simplistic, comfortable, easy & 0000 / 123987 & ``To just chose an easy PIN.'' \\
    System & \,~8     & User's established systematic strategy & 0433 / 041512 & ``I used the numbers from the current time 04:33 PM.'' \\
    Phone & \,~3     & (Partial) phone number & 1601 / 407437 & ``I used the first four digits of a friend's phone number.'' \\
    \bottomrule
    \end{tabular}}
\end{table}

\vspace{-1em}
\begin{table}[htbp]
  \caption{We coded and analyzed a sample of 126~PIN changing strategies of participants that encountered a blocklist and in response changed their PIN. Below we list and explain our codes. Two coders independently coded the data. The level of agreement among the coders, measured by Cohen's kappa was $\kappa = 0.96$. Question: ``\emph{After receiving the warning message, please describe how or if your strategy changed when choosing your PIN.}''}\label{tab:changing-strategies-codebook}
  \centering
    \resizebox{0.9\columnwidth}{!}{\begin{tabular}{ccllll}
    \toprule
    \multicolumn{1}{l}{\textbf{Code Name}} & \textbf{Frequency} & \textbf{Description} & \textbf{Use Case} & \textbf{Strategy} & \textbf{Sample from the Study} \\
    \midrule
    \multirow{2}[0]{*}{Same} & \multirow{2}[0]{*}{28} & \multirow{2}[0]{*}{Same strategy for both} & Selection & Date  & ``Birthday of relative.'' \\
          & & & Change & Date  & ``Chose another birthday.'' \\
    \midrule
    \multirow{2}[0]{*}{Minor} & \multirow{2}[0]{*}{33} & \multirow{2}[0]{*}{Slight modification of strategy} & Selection & Meaning & ``It's one I remember, a number with personal significance.'' \\
          & & & Change & Meaning++ & ``I changed one number in the sequence to get the app to accept it.'' \\
    \midrule
    \multirow{2}[0]{*}{New} & \multirow{2}[0]{*}{65} & \multirow{2}[0]{*}{New strategy that is different} & Selection & Date  & ``I used my girlfriend's birthday.'' \\
          & & & Change & Phone & ``I changed my strategy to a memorable phone number's last 4 digits.'' \\
    \bottomrule
    \end{tabular}}
\end{table} 

%% file: main.bbl
\begin{thebibliography}{10}

\bibitem{akhawe-13-alice}
Devdatta Akhawe and Adrienne~Porter Felt.
\newblock {Alice in Warningland: A Large-Scale Field Study of Browser Security
  Warning Effectiveness}.
\newblock In {\em USENIX Security Symposium}, SSYM~'13, pages 257--272,
  Washington, District of Columbia, USA, July 2013. USENIX.

\bibitem{amitay-11-iphone-pins}
Daniel Amitay.
\newblock {Most Common iPhone Passcodes}, June 2011.
\newblock
  \url{http://danielamitay.com/blog/2011/6/13/most-common-iphone-passcodes}, as
  of \today.

\bibitem{android-18-masterkey}
{Android Open Source Project}.
\newblock {Full-Disk Encryption -- Storing the Encrypted Key}, August 2018.
\newblock
  \url{https://source.android.com/security/encryption/full-disk\#storing_the_encrypted_key},
  as of \today.

\bibitem{android-19-gatekeeper}
{Android Open Source Project}.
\newblock {Android 10: GateKeeper -- ComputeRetryTimeout Function}, September
  2019.
\newblock
  \url{https://android.googlesource.com/platform/system/gatekeeper/+/refs/heads/android10-release/gatekeeper.cpp\#261},
  as of \today.

\bibitem{apple-19-ios-security}
{Apple, Inc.}
\newblock {Apple Platform Security}, December 2019.
\newblock
  \url{https://manuals.info.apple.com/MANUALS/1000/MA1902/en_US/apple-platform-security-guide.pdf},
  as of \today.

\bibitem{aviv-15-pattern-bigger}
Adam~J. Aviv, Devon Budzitowski, and Ravi Kuber.
\newblock {Is Bigger Better? Comparing User-Generated Passwords on 3x3 vs. 4x4
  Grid Sizes for Android's Pattern Unlock}.
\newblock In {\em Annual Computer Security Applications Conference}, ACSAC~'15,
  pages 301--310, Los Angeles, California, USA, December 2015. ACM.

\bibitem{aviv-17-shoulder-surfing-baseline}
Adam~J. Aviv, John~T. Davin, Flynn Wolf, and Ravi Kuber.
\newblock {Towards Baselines for Shoulder Surfing on Mobile Authentication}.
\newblock In {\em Annual Conference on Computer Security Applications},
  ACSAC~'17, pages 486--498, Orlando, Florida, USA, December 2017. ACM.

\bibitem{aviv-10-smudge}
Adam~J. Aviv, Katherine Gibson, Evan Mossop, Matt Blaze, and Jonathan~M. Smith.
\newblock {Smudge Attacks on Smartphone Touch Screens}.
\newblock In {\em USENIX Workshop on Offensive Technologies}, WOOT~'10, pages
  1--7, Washington, District of Columbia, USA, August 2010. USENIX.

\bibitem{aviv-18-comparing-shoulder-surfing}
Adam~J. Aviv, Flynn Wolf, and Ravi Kuber.
\newblock {Comparing Video Based Shoulder Surfing with Live Simulation and
  Towards Baselines for Shoulder Surfing on Mobile Authentication}.
\newblock In {\em Annual Conference on Computer Security Applications},
  ACSAC~'18, pages 486--498, Puerto Rico, USA, December 2018. ACM.

\bibitem{bonneau-12-entropy}
Joseph Bonneau.
\newblock {The Science of Guessing: Analyzing an Anonymized Corpus of 70
  Million Passwords}.
\newblock In {\em IEEE Symposium on Security and Privacy}, SP~'12, pages
  538--552, San Jose, California, USA, May 2012. IEEE.

\bibitem{bonneau-12-pin}
Joseph Bonneau, S{\"o}ren Preibusch, and Ross Anderson.
\newblock {A Birthday Present Every Eleven Wallets? The Security of
  Customer-Chosen Banking PINs}.
\newblock In {\em Financial Cryptography and Data Security}, FC~'12, pages
  25--40, Kralendijk, Bonaire, February 2012. Springer.

\bibitem{brewster-18-graykey}
Thomas Brewster.
\newblock {Mysterious \$15,000 ``GrayKey'' Promises To Unlock iPhone X For The
  Feds}, March 2018.
\newblock
  \url{https://www.forbes.com/sites/thomasbrewster/2018/03/05/apple-iphone-x-graykey-hack/},
  as of \today.

\bibitem{brewster-18-cellebrite}
Thomas Brewster.
\newblock {The Feds Can Now (Probably) Unlock Every iPhone Model In Existence},
  February 2018.
\newblock
  \url{https://www.forbes.com/sites/thomasbrewster/2018/02/26/government-can-access-any-apple-iphone-cellebrite/},
  as of \today.

\bibitem{cherapau-15-biometrics-passcodes}
Ivan Cherapau, Ildar Muslukhov, Nalin Asanka, and Konstantin Beznosov.
\newblock {On the Impact of Touch ID on iPhone Passcodes}.
\newblock In {\em Symposium on Usable Privacy and Security}, SOUPS~'15, pages
  257--276, Ottawa, Canada, July 2015. USENIX.

\bibitem{felt-15-ssl-warnings}
Adrienne~Porter Felt, Alex Ainslie, Robert~W. Reeder, Sunny Consolvo, Somas
  Thyagaraja, Alan Bettes, Helen Harris, and Jeff Grimes.
\newblock {Improving SSL Warnings: Comprehension and Adherence}.
\newblock In {\em ACM Conference on Human Factors in Computing Systems},
  CHI~'15, pages 2893--2902, Seoul, Republic of Korea, April 2015. ACM.

\bibitem{golla-18-psm}
Maximilian Golla and Markus D\"{u}rmuth.
\newblock {On the Accuracy of Password Strength Meters}.
\newblock In {\em ACM Conference on Computer and Communications Security},
  CCS~'18, pages 1567--1582, Toronto, Ontario, Canada, October 2018. ACM.

\bibitem{golla-19-pattern-meter-wip}
Maximilian Golla, Jan Rimkus, Adam~J. Aviv, and Markus D\"{u}rmuth.
\newblock {Work in Progress: On the In-Accuracy and Influence of Android
  Pattern Strength Meters}.
\newblock In {\em Workshop on Usable Security and Privacy}, USEC~'19, San
  Diego, California, USA, February 2019. ISOC.

\bibitem{golla-18-what-was}
Maximilian Golla, Miranda Wei, Juliette Hainline, Lydia Filipe, Markus
  D\"{u}rmuth, Elissa Redmiles, and Blase Ur.
\newblock {``What was that site doing with my Facebook password?'' Designing
  Password-Reuse Notification}.
\newblock In {\em ACM Conference on Computer and Communications Security},
  CCS~'18, pages 1549--1566, Toronto, Canada, November 2018. ACM.

\bibitem{gosney-16-linkedin-sha1}
Jeremi~M. Gosney~(``epixoip'').
\newblock {How LinkedIn's Password Sloppiness Hurts Us All}, June 2016.
\newblock \url{https://arstechnica.com/?post_type=post&p=892339}, as of \today.

\bibitem{nist-17-sp800-63b}
Paul~A. Grassi, James~L. Fenton, and William~E. Burr.
\newblock {Digital Identity Guidelines -- Authentication and Lifecycle
  Management: NIST Special Publication 800-63B}, June 2017.

\bibitem{greene-14-cant-type-that}
Kristen~K. Greene, Melissa~A. Gallagher, Brian~C. Stanton, and Paul~Y. Lee.
\newblock {I Can't Type That! P@\$\$w0rd Entry on Mobile Devices}.
\newblock In {\em Human Aspects of Information Security, Privacy, and Trust},
  HAS~'14, pages 160--171, Heraklion, Greece, June 2014. Springer.

\bibitem{harbach-14-hard-lock-life}
Marian Harbach, Emanuel {von Zezschwitz}, Andreas Fichtner, Alexander {De
  Luca}, and Matthew Smith.
\newblock {It's a Hard Lock Life: A Field Study of Smartphone (Un)Locking
  Behavior and Risk Perception}.
\newblock In {\em Symposium on Usable Privacy and Security}, SOUPS~'14, pages
  213--230, Menlo Park, California, USA, July 2014. USENIX.

\bibitem{hunt-18-500m}
Troy Hunt.
\newblock {I've Just Launched ``Pwned Passwords'' V2 With Half a Billion
  Passwords for Download}, February 2018.
\newblock
  \url{https://www.troyhunt.com/ive-just-launched-pwned-passwords-version-2/},
  as of \today.

\bibitem{kelley-12-again}
Patrick Kelley, Saranga Kom, Michelle~L. Mazurek, et~al.
\newblock {Guess Again (and Again and Again): Measuring Password Strength by
  Simulating Password-Cracking Algorithms}.
\newblock In {\em IEEE Symposium on Security and Privacy}, SP~'12, pages
  523--537, San Jose, California, USA, May 2012. IEEE.

\bibitem{kim-12-pin-policies}
Hyoungshick Kim and Jun~Ho Huh.
\newblock {PIN Selection Policies: Are They Really Effective?}
\newblock {\em Computers \& Security}, 31(4):484--496, June 2012.

\bibitem{lee-97-tough}
Fiona Lee.
\newblock {When the Going Gets Tough, Do the Tough Ask for Help? Help Seeking
  and Power Motivation in Organizations}.
\newblock {\em Organizational Behavior and Human Decision Processes},
  72(3):336--363, December 1997.

\bibitem{loge-16-pattern-user-choice}
Marte L{\o}ge, Markus D\"urmuth, and Lillian R{\o}stad.
\newblock {On User Choice for Android Unlock Patterns}.
\newblock In {\em European Workshop on Usable Security}, EuroUSEC~'16,
  Darmstadt, Germany, July 2016. ISOC.

\bibitem{melicher-16-smartphone-passwords}
William Melicher, Darya Kurilova, Sean~M. Segreti, Pranshu Kalvani, Richard
  Shay, Blase Ur, Lujo Bauer, Nicolas Christin, Lorrie~Faith Cranor, and
  Michelle~L. Mazurek.
\newblock {Usability and Security of Text Passwords on Mobile Devices}.
\newblock In {\em ACM Conference on Human Factors in Computing Systems},
  CHI~'16, pages 527--539, San Jose, CA, USA, May 2016. ACM.

\bibitem{mohammad-13-nrc-database}
Saif~M. Mohammad and Peter~D. Turney.
\newblock {Crowdsourcing a Word-Emotion Association Lexicon}.
\newblock {\em Computational Intelligence}, 29(3):436--465, 2013.

\bibitem{newman-19-encrypt-cheap-phones}
Lily~Hay Newman.
\newblock {Google's Making it Easier to Encrypt Even Cheap Android Phones},
  February 2019.
\newblock
  \url{https://www.wired.com/story/android-encryption-cheap-smartphones/}, as
  of \today.

\bibitem{qiu-19-age-effects}
Lina Qiu, Alexander De~Luca, Ildar Muslukhov, and Konstantin Beznosov.
\newblock {Towards Understanding the Link Between Age and Smartphone
  Authentication}.
\newblock In {\em ACM Conference on Human Factors in Computing Systems},
  CHI~'19, pages 163:1--163:10, Glasgow, Scotland, United Kingdom, May 2019.
  ACM.

\bibitem{redmiles-17-best-practices}
Elissa~M. Redmiles, Yasemin Acar, Sascha Fahl, and Michelle~L. Mazurek.
\newblock {A Summary of Survey Methodology Best Practices for Security and
  Privacy Researchers}.
\newblock Technical Report CS-TR-5055, UM Computer Science Department, May
  2017.

\bibitem{reed-18-graykey-malwarebytes}
Thomas Reed.
\newblock {GrayKey iPhone Unlocker Poses Serious Security Concerns}, March
  2018.
\newblock \url{https://blog.malwarebytes.com/?p=22342}, as of \today.

\bibitem{renaud-15-mental-models}
Karen Renaud and Melanie Volkamer.
\newblock {Exploring Mental Models Underlying PIN Management Strategies}.
\newblock In {\em World Congress on Internet Security}, WorldCIS~'15, pages
  19--21, Dublin, United Kingdom, October 2015. IEEE.

\bibitem{schaub-12-passwords-on-smartphones}
Florian Schaub, Ruben Deyhle, and Michael Weber.
\newblock {Password Entry Usability and Shoulder Surfing Susceptibility on
  Different Smartphone Platforms}.
\newblock In {\em International Conference on Mobile and Ubiquitous
  Multimedia}, MUM~'12, pages 13:1--13:10, Ulm, Germany, December 2012. ACM.

\bibitem{shay-15-sugar}
Richard Shay, Lujo Bauer, Nicolas Christin, Lorrie~Faith Cranor, Alain Forget,
  Saranga Komanduri, Michelle~L. Mazurek, William Melicher, Sean~M. Segreti,
  and Blase Ur.
\newblock {A Spoonful of Sugar?: The Impact of Guidance and Feedback on
  Password-Creation Behavior}.
\newblock In {\em ACM Conference on Human Factors in Computing Systems},
  CHI~'15, pages 2903--2912, Seoul, Republic of Korea, April 2015. ACM.

\bibitem{stark-19-urlephant}
Emily Stark.
\newblock {The URLephant}.
\newblock In {\em USENIX Enigma Conference}, Enigma~'19, Burlingame,
  California, USA, January 2019. USENIX.

\bibitem{sunshine-09-crying-wolf}
Joshua Sunshine, Serge Egelman, Hazim Almuhimedi, Neha Atri, and Lorrie~Faith
  Cranor.
\newblock {Crying Wolf: An Empirical Study of SSL Warning Effectiveness}.
\newblock In {\em USENIX Security Symposium}, SSYM~'09, pages 399--416, San
  Diego, California, USA, June 2009. USENIX.

\bibitem{uellenbeck-13-pattern}
Sebastian Uellenbeck, Markus D\"urmuth, Christopher Wolf, and Thorsten Holz.
\newblock {Quantifying the Security of Graphical Passwords: The Case of Android
  Unlock Patterns}.
\newblock In {\em ACM Conference on Computer and Communications Security},
  CCS~'13, pages 161--172, Berlin, Germany, November 2016. ACM.

\bibitem{ur-17-data-driven-pm}
Blase Ur, Felicia Alfieri, Maung Aung, Lujo Bauer, Nicolas Christin, Jessica
  Colnago, Lorrie~Faith Cranor, Henry Dixon, Pardis~Emami Naeini, Hana Habib,
  Noah Johnson, and William Melicher.
\newblock {Design and Evaluation of a Data-Driven Password Meter}.
\newblock In {\em ACM Conference on Human Factors in Computing Systems},
  CHI~'17, pages 3775--3786, Denver, Colorado, USA, May 2017. ACM.

\bibitem{ur-15-added-at-the-end}
Blase Ur, Fumiko Noma, Jonathan Bees, Sean~M. Segreti, Richard Shay, Lujo
  Bauer, Nicolas Christin, and Lorrie~Faith Cranor.
\newblock {``I Added `!' at the End to Make It Secure'': Observing Password
  Creation in the Lab}.
\newblock In {\em Symposium on Usable Privacy and Security}, SOUPS~'15, pages
  123--140, Ottawa, Ontario, Canada, July 2015. USENIX.

\bibitem{ur-15-pgs}
Blase Ur, Sean~M. Segreti, Lujo Bauer, Nicolas Christin, Lorrie~Faith Cranor,
  Saranga Komanduri, Darya Kurilova, Michelle~L. Mazurek, William Melicher, and
  Richard Shay.
\newblock {Measuring Real-World Accuracies and Biases in Modeling Password
  Guessability}.
\newblock In {\em USENIX Security Symposium}, SSYM~'15, pages 463--481,
  Washington, District of Columbia, USA, August 2015. USENIX.

\bibitem{dhs-12-menlo-report}
{U.S. Department of Homeland Security}.
\newblock {The Menlo Report: Ethical Principles Guiding Information and
  Communication Technology Research}, August 2012.
\newblock
  \url{https://www.caida.org/publications/papers/2012/menlo_report_actual_formatted/},
  as of \today.

\bibitem{zezschwitz-14-honey-i-shrunk}
Emanuel {von Zezschwitz}, Alexander De~Luca, and Heinrich Hussmann.
\newblock {Honey, I Shrunk the Keys: Influences of Mobile Devices on Password
  Composition and Authentication Performance}.
\newblock In {\em Nordic Conference on Human-Computer Interaction},
  NordiCHI~'14, pages 461--470, Helsinki, Finland, October 2014. ACM.

\bibitem{zezschwitz-16-pattern-effective-space}
Emanuel {von Zezschwitz}, Malin Eiband, Daniel Buschek, Sascha Oberhuber,
  Alexander De~Luca, Florian Alt, and Heinrich Hussmann.
\newblock {On Quantifying the Effective Passsword Space of Grid-Based Unlock
  Gestures}.
\newblock In {\em Conference on Mobile and Ubiquitous Multimedia}, MUM~'16,
  pages 201--212, Rovaniemi, Finland, December 2016. ACM.

\bibitem{wang-17-pin}
Ding Wang, Qianchen Gu, Xinyi Huang, and Ping Wang.
\newblock {Understanding Human-Chosen PINs: Characteristics, Distribution and
  Security}.
\newblock In {\em ACM Asia Conference on Computer and Communications Security},
  ASIA~CCS~'17, pages 372--385, Abu Dhabi, United Arab Emirates, April 2017.
  ACM.

\bibitem{welch-18-ios-usb-restricted-mode}
Chris Welch.
\newblock {Apple Releases iOS 11.4.1 and Blocks Passcode Cracking Tools Used by
  Police}, July 2018.
\newblock \url{https://www.theverge.com/2018/7/9/17549538/}, as of \today.

\bibitem{wichmann-11-selfdetermintation}
Sonia~Secher Wichmann.
\newblock {Self-Determination Theory: The Importance of Autonomy to Well-Being
  Across Cultures}.
\newblock {\em Journal of Humanistic Counseling}, 50(1):16--26, March 2011.

\bibitem{yang-14-entry-method-affects}
Yulong Yang, Janne Lindqvist, and Antti Oulasvirta.
\newblock {Text Entry Method Affects Password Security}.
\newblock In {\em Learning from Authoritative Security Experiment Results},
  LASER~'14, pages 11--20, Arlington, Virginia, USA, October 2014. USENIX.

\end{thebibliography}
